\documentclass[twocolumn,amsmath,amssymb,prapplied,longbibliography]{revtex4}

\usepackage{graphicx}
\usepackage{dcolumn}
\usepackage{bm}
\usepackage{color}
\usepackage{notes2bib}

\begin{document}



\title{Local vibrational modes of Si vacancy spin qubits in SiC}

\author{Z.~Shang$^{1}$}
\author{A.~Hashemi$^{2}$}
\author{Y.~Berenc{\'e}n$^{1}$}
\author{H.-P.~Komsa$^{2,3}$}
\author{P.~Erhart$^{4}$}
\author{A.~V.~Krasheninnikov$^{1,2}$}
\author{G.~V.~Astakhov$^{1}$}
\email[E-mail:~]{g.astakhov@hzdr.de}

\affiliation{$^1$Institute of Ion Beam Physics and Materials Research, Helmholtz-Zentrum Dresden-Rossendorf, Dresden, Germany  \\
$^2$Department of Applied Physics, Aalto University, Espoo, Finland.\\
$^3$Microelectronics Research Unit, University of Oulu, Oulu, Finland. \\
$^4$Department of Physics, Chalmers University of Technology, Gothenburg, Sweden.
 }

\begin{abstract}
Silicon carbide is a very promising platform  for quantum applications because of extraordinary spin and optical properties of point defects in this technologically-friendly material. These properties are strongly influenced by crystal vibrations, but the exact relationship between them and the behavior of spin qubits is not fully investigated. We uncover the local vibrational modes of the Si vacancy spin qubits in as-grown 4H-SiC. We apply the resonant microwave field to isolate the contribution from one particular type of defects, the so-called V2 center, and observe the zero-phonon line together with seven equally-separated phonon replicas. Furthermore, we present first-principles calculations of the photoluminescence lineshape, which are in excellent agreement with our experimental data. To boost up the calculation accuracy and decrease the computation time, we extract the force constants using machine learning algorithms. This allows us to identify dominant modes in the lattice vibrations coupled to an excited electron during optical emission in the Si vacancy.  The resonance phonon energy of $36 \, \mathrm{meV}$ and the Debye-Waller factor of about $6\%$ are obtained. We establish experimentally that the activation energy of the optically-induced spin polarization is given by the local vibrational energy. Our findings give insight into the coupling of electronic states to vibrational modes in SiC spin qubits, which is essential to predict their spin, optical, mechanical and thermal properties. The approach described can be applied to a large variety of spin defects with spectrally overlapped contributions in SiC as well as in other 3D and 2D  materials.
\end{abstract}

 

\maketitle

\section*{Introduction}

Since the demonstration of promising quantum properties of intrinsic point defects in silicon carbide (SiC) \cite{Baranov:2011ib, Koehl:2011fv, Riedel:2012jq, Soltamov:2012ey}, they have been used to implement room-temperature quantum emitters \cite{Kraus:2013di, Castelletto:2013el, Christle:2014ti, Widmann:2014ve, Fuchs:2015ii} as well as to realize quantum sensing of magnetic fields \cite{Kraus:2013vf, Simin:2015dn, Simin:2016cp, Niethammer:2016bc, Cochrane:2016dd, Soykal:2016tk, Soltamov:2019hr}, electric fields \cite{Wolfowicz:2019cz} and temperature \cite{Kraus:2013vf, Anisimov:2016er, Zhou:2017cca}. Particularly, silicon vacancies ($\mathrm{V_{Si}}$) and silicon-carbon divacancies ($\mathrm{VV}$) in SiC reveal extremely long spin coherence time \cite{Yang:2014kqa, Christle:2014ti, Widmann:2014ve, Carter:2015vc, Seo:2016ey, Embley:2017bf, Simin:2017iw, Fischer:2018fj, Soltamov:2019hr} and hold promise to implement quantum repeaters due to inherent spin-photon interface and high spectral stability \cite{Economou:2016bp, Christle:2017tq, Nagy:2019fw, Udvarhelyi:2019eh}. Existing device fabrication protocols on the wafer scale in combination with  3D defect engineering \cite{Kraus:2017cka, Wang:2017fb, Wang:2019dk} allow manufacturing integrated quantum devices \cite{Fuchs:2013dz, Lohrmann:2015hd, Sato:2018jq, Widmann:2018jh} with electrical \cite{Klimov:2013ua, Falk:2014fh, Widmann:2019ja} and mechanical \cite{Falk:2014fh, Whiteley:2019eu, Poshakinskiy:2019bi} control of defect spin qubits. SiC nanocrystals with color centers are also suggested as fluorescence biomarkers in biomedical applications \cite{Castelletto:2014eu, Muzha:2014th}.   

Vacancies can be imagined as artificial atoms incorporated into SiC lattice. The communication with them is usually realized through optical excitation and photoluminescence (PL) detection.  A fingerprint of each defect at cryogenic temperatures is the spectrally narrow emission at certain wavelength, the so-called zero-phonon line (ZPL)  \cite{Sorman:2000ij, Son:2006im, Baranov:2011ib, Fuchs:2015ii}. Unlike atoms, radiative recombination in point defects is accompanied by phonon emission due to  the interaction with lattice vibrations. This process results  in the phonon side band (PSB), which is spectrally shifted towards longer wavelength relative to the ZPL \cite{Gali:2011fn, Alkauskas:2014gq}. A high ratio of the emitted light from the ZPL to the all emitted light, the Debye-Waller (DW) factor, is necessary for the implementation of quantum repeaters. The local vibrational energy also contributes to the spin-lattice relaxation time $T_1$ \cite{Simin:2017iw}. 

Although the understanding of PSB is important for quantum applications, it has not been investigated systematically in SiC. The previous works \cite{Nagy:2018ey, Banks:2019je} are limited to the report of the upper limit for the DW factor in a single $\mathrm{V_{Si}}$ defect, which is below 30\%-40\% depending on the crystallographic site and polytype.  Most of the theoretical works are concentrated on the spin-optical properties \cite{Ivady:2018cu, Bockstedte:2018dh}  and no first-principles calculation of the PL lineshape in SiC has been performed to date. 

In this work, we present the measurement of the V2 $\mathrm{V_{Si}}$ PL spectrum in polytype 4H-SiC,  consisting of ZPL and 7 increasingly broad phonon replicas. We use optically detected magnetic resonance (ODMR)  to clearly separate spectrally overlapped contributions from other $\mathrm{V_{Si}}$ and $\mathrm{VV}$ defects. This approach allows us to unambiguously determine the V2 local vibrational energy as the separation between two adjacent PSB peaks and the DW factor. 
We also find the activation energy of the spin polarization from the PSB temperature evolution. To shed more light into the microscopic nature of the PSB, we calculate the lineshape using density functional theory (DFT). The experimental lineshape is very well reproduced and concurrently also leads to close values for DW factor and the local vibrational energy.
The electron-phonon coupling spectral density, partially phonon-contributed lineshapes, and unfolded phonon dispersion curves are then used to 
explore and visualize the origin of dominant lattice vibrations.

\section*{Experiment}

The PL and ODMR measurements are carried out on a homemade setup. The samples are excited with an 808-nm laser, which is modulated by a chopper at $20 \, \mathrm{Hz}$.  The PL signal is collected by a lens group and focused onto the entrance slit of a monochromator. 
An InGaAs detector converts the PL intensity into a photovoltage, which is amplified and read out by a lock-in amplifier. For the ODMR measurements, the chopper is removed and a commercial signal generator provides microwaves (MW) modulated at $20 \, \mathrm{Hz}$. The MW field guided into a coplanar waveguide induces spin transitions in the sample placed on the top of the waveguide.  The MW transmission loses at the measuring point of the sample are less than $0.1 \, \mathrm{dB}$ in the scanning MW frequency range of $40 \, \mathrm{MHz}$ to $100 \, \mathrm{MHz}$. The modulated $\mathrm{\Delta PL}$ signal is collected by the same lens group and read out by the lock-in amplifier. 


\begin{figure}[t]
\includegraphics[width=.44\textwidth]{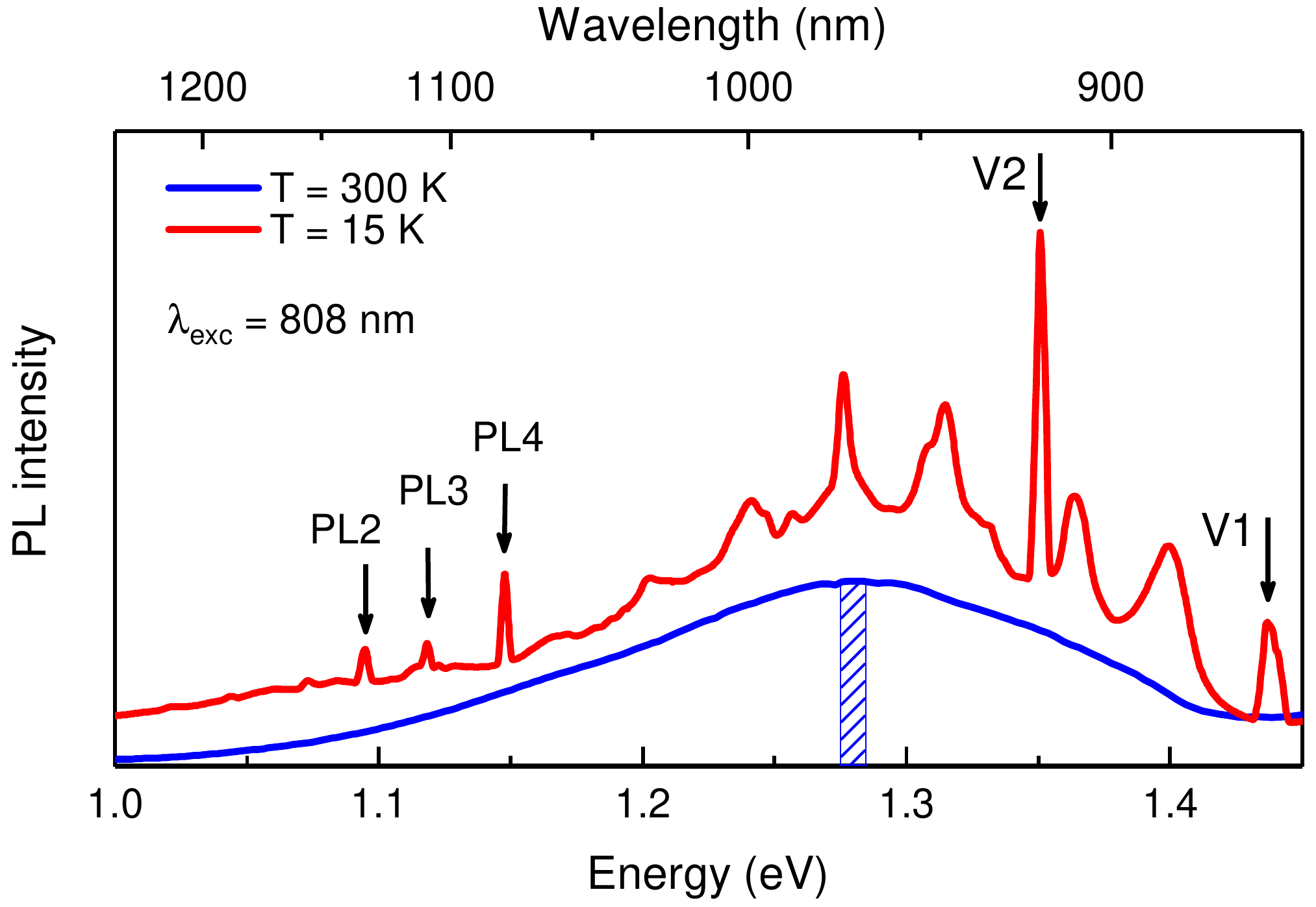}
\caption{Room-temperature and low-temperature PL from $\mathrm{V_{Si}}$ in pristine 4H-SiC. 
At $T = 300 \, \mathrm{K}$, the shadow area around $970 \, \mathrm{nm}$  indicates the spectral resolution $\Delta \lambda = 5.5 \, \mathrm{nm}$. At $T = 15 \, \mathrm{K}$, the zero phonon lines of two distinct $\mathrm{V_{Si}}$ centers, V1 (ZPL at $863 \, \mathrm{nm}$) and V2 (ZPL at $918 \, \mathrm{nm}$), are clearly observed. The spectral resolution is  $\Delta \lambda = 1.2 \, \mathrm{nm}$ ($\approx$ 2 meV). } \label{fig1}
\end{figure}

The sample under study is a piece diced from a pristine high-purity semi-insulating (HPSI) 4H-SiC wafer purchased from Cree. 
It is not irradiated and contains native $\mathrm{V_{Si}}$ and $\mathrm{VV}$  defects.  The sample is mounted on the cold finger of a closed-cycle cryostat and the experiments are performed in the temperature range from $T = 300 \, \mathrm{K}$ down to $T = 15 \, \mathrm{K}$. To increase the PL intensity at $T = 300 \, \mathrm{K}$, we use a relatively wide monochromator entrance slit of $2 \, \mathrm{mm}$ with corresponding spectral resolution of $5.5 \, \mathrm{nm}$. At $T = 15 \, \mathrm{K}$,  the entrance slit size is reduced to $0.5 \, \mathrm{mm}$ to improve the spectral resolution to $1.3 \, \mathrm{nm}$. 

Typical PL spectra of the sample under study are presented in Fig.~\ref{fig1}. A wide emission band with the maximum at around $1.28 \, \mathrm{eV}$ ($970 \, \mathrm{nm}$), associated with the $\mathrm{V_{Si}}$ defects  \cite{Sorman:2000ij}, is clearly observed at $T = 300 \, \mathrm{K}$. The emission band transfers to a series of ZPLs when the sample is cooled down to $T = 15 \, \mathrm{K}$. Two ZPLs at $1.44 \, \mathrm{eV}$ ($863 \, \mathrm{nm}$) and $1.35 \, \mathrm{eV}$ ($918 \, \mathrm{nm}$) correspond to the V1 and V2 $\mathrm{V_{Si}}$ defects, respectively \cite{Sorman:2000ij, Fuchs:2015ii}. Several other ZPLs (labeled as PL2-PL4) are observed in the spectral range $1.1-1.2 \, \mathrm{eV}$ and related to the silicon-carbon $\mathrm{VV}$ defects  \cite{Son:2006im}. 

\begin{figure}[t]
\includegraphics[width=.48\textwidth]{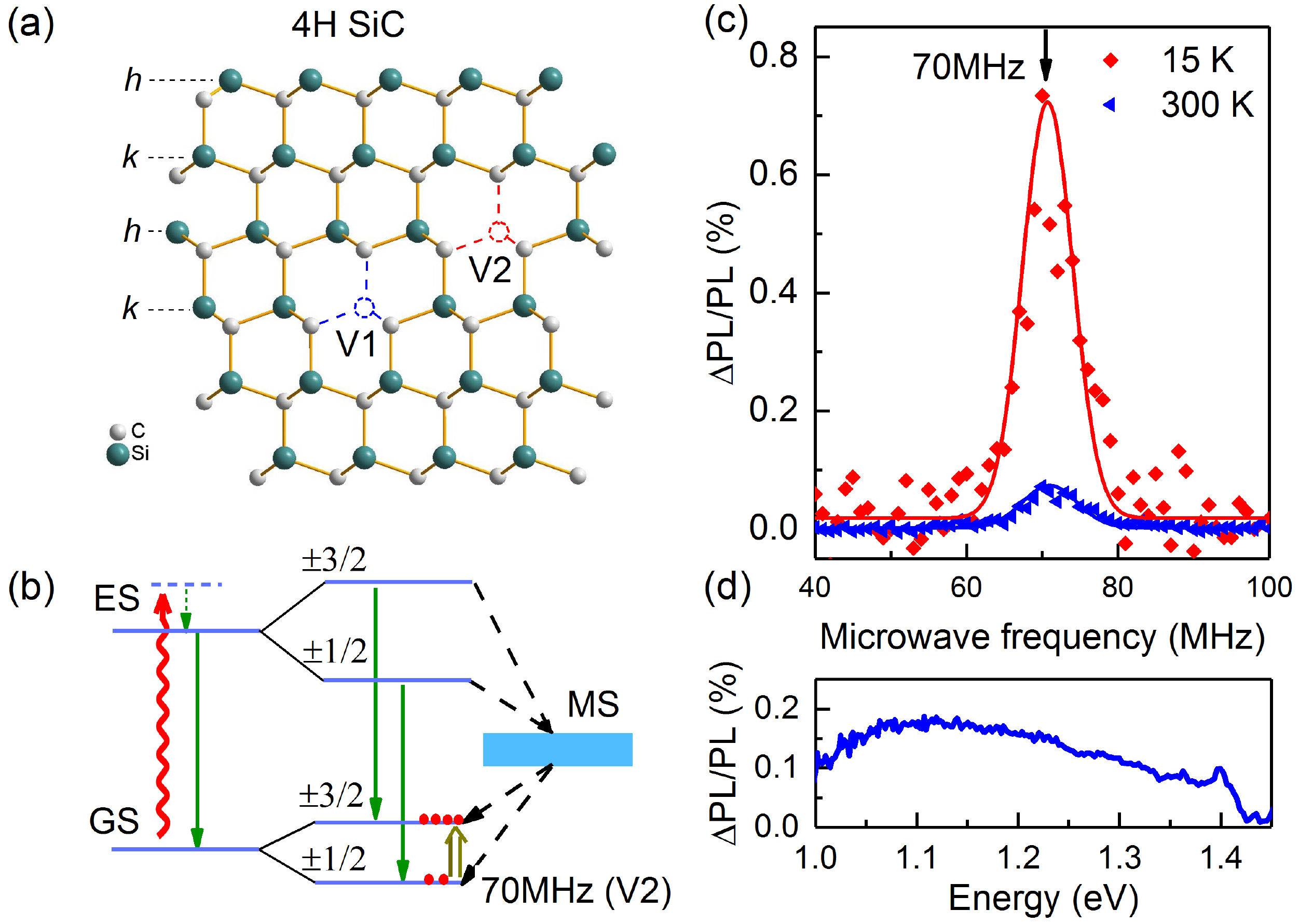}
\caption{(a) Schematic representation of the V1 and V2 $\mathrm{V_{Si}}$ defects associated with different crystallographic sites in 4H-SiC. (b) A scheme of the V2 spin pumping process. The red line presents the V2 spin center excited by an $808 \, \mathrm{nm}$ laser. It relaxes to the ground state by two process: radiative recombination (solid lines)  and spin dependent relaxation through the metastable states (MS)  (dashed lines). The double arrow shows spin manipulation by the MW field at $70 \, \mathrm{MHz}$. (c) The V2 ODMR spectrum at room temperature and low temperature. The arrow indicates the resonance frequency of $70 \, \mathrm{MHz}$ at zero magnetic field. (d) The  V2 ODMR contrast at $70 \, \mathrm{MHz}$ as a function of detection energy over the PL spectrum at room temperature. } \label{fig2}
\end{figure}

We concentrate on the V2 $\mathrm{V_{Si}}$ defect, associated with one of the two possible crystallographic sites in 4H-SiC (Fig.~\ref{fig2}(a)). The mechanism of the zero-field ODMR associated with the $\mathrm{V_{Si}}$ defects is qualitatively explained in Fig.~\ref{fig2}(b). The $\mathrm{V_{Si}}$ has spin $S = 3/2$ in the ground state (GS) and excited state (ES) \cite{Kraus:2013di}.  Optical excitation of the V2  $\mathrm{V_{Si}}$ defect into the ES is followed by two processes, radiative recombination to the ground state GS (solid lines) and non-radiative spin-dependent relaxation (dashed lines) through the metastable state (MS). Application of the resonant MW field at $70 \, \mathrm{MHz}$, which is equal to the zero-field splitting between the $m_S = \pm 1/2$ and $m_S = \pm 3/2$ states, changes the population of these spin sublevels. It breaks the equilibrium between the relaxation processes resulting in non-zero $\mathrm{ \Delta PL}$ \cite{Kraus:2013vf}. 

Figure~\ref{fig2}(c)  presents the ODMR contrast ($\mathrm{ \Delta PL / PL}$) as a function of MW frequency. The PL is detected at $970 \, \mathrm{nm}$ at $T = 300 \, \mathrm{K}$ (shadow area in Fig.~\ref{fig1}) and at the V2 ZPL at $T = 15 \, \mathrm{K}$. To ensure that ODMR experiments are performed under optimum conditions, we investigate the laser power and MW power dependences at different temperatures. The ODMR contrast saturates in both cases  \cite{Fischer:2018fj} and  we obtain  $C_{max} = 0.80 \pm 0.02 \%$ and $C_{max} = 0.21 \pm 0.05 \%$ for $T = 15 \, \mathrm{K}$ and $T = 300 \, \mathrm{K}$, respectively (Fig.~\ref{fig2}(c)). Remarkably, the ODMR contrast only marginally depends on the detection energy over the PL band ($1.1 - 1.35 \, \mathrm{eV}$) at room temperature (Fig.~\ref{fig2}(d)).  

\begin{figure}[t]
\includegraphics[width=.48\textwidth]{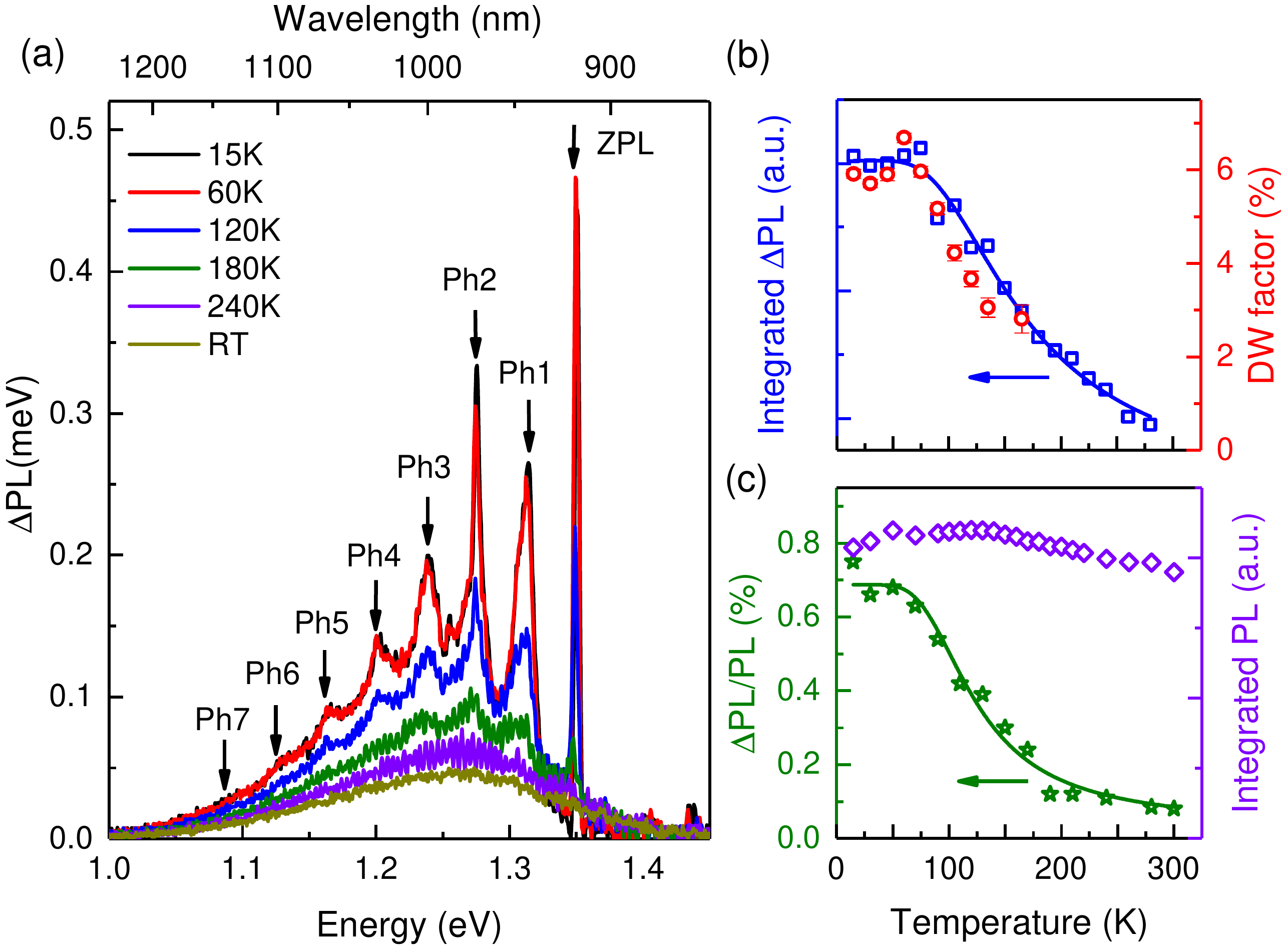}
\caption{(a) Evolution of the $\mathrm{ \Delta PL}$  spectrum with temperature under applied MW field at $70 \, \mathrm{MHz}$.  (b) Temperature dependence of the spectrally integrated V2 $\mathrm{ \Delta PL}$ and the DW factor. The solid line is a fit to Eq.~(\ref{Arrh}) with the activation energy $E_A = 39 \pm 4 \, \mathrm{meV}$. (c) Temperature dependence of the ODMR contrast $\mathrm{ \Delta PL / PL}$ detected at the V2 ZPL and the integrated PL. The solid line is a fit to Eq.~(\ref{Arrh}) with the activation energy $E_A = 39 \pm 3 \, \mathrm{meV}$ and  $\mathrm{ \Delta PL} (0)$ replaced with  $\mathrm{ \Delta PL / PL} (0)$. } \label{fig3}
\end{figure}

\begin{figure*}[t]
\includegraphics[width=.73\textwidth]{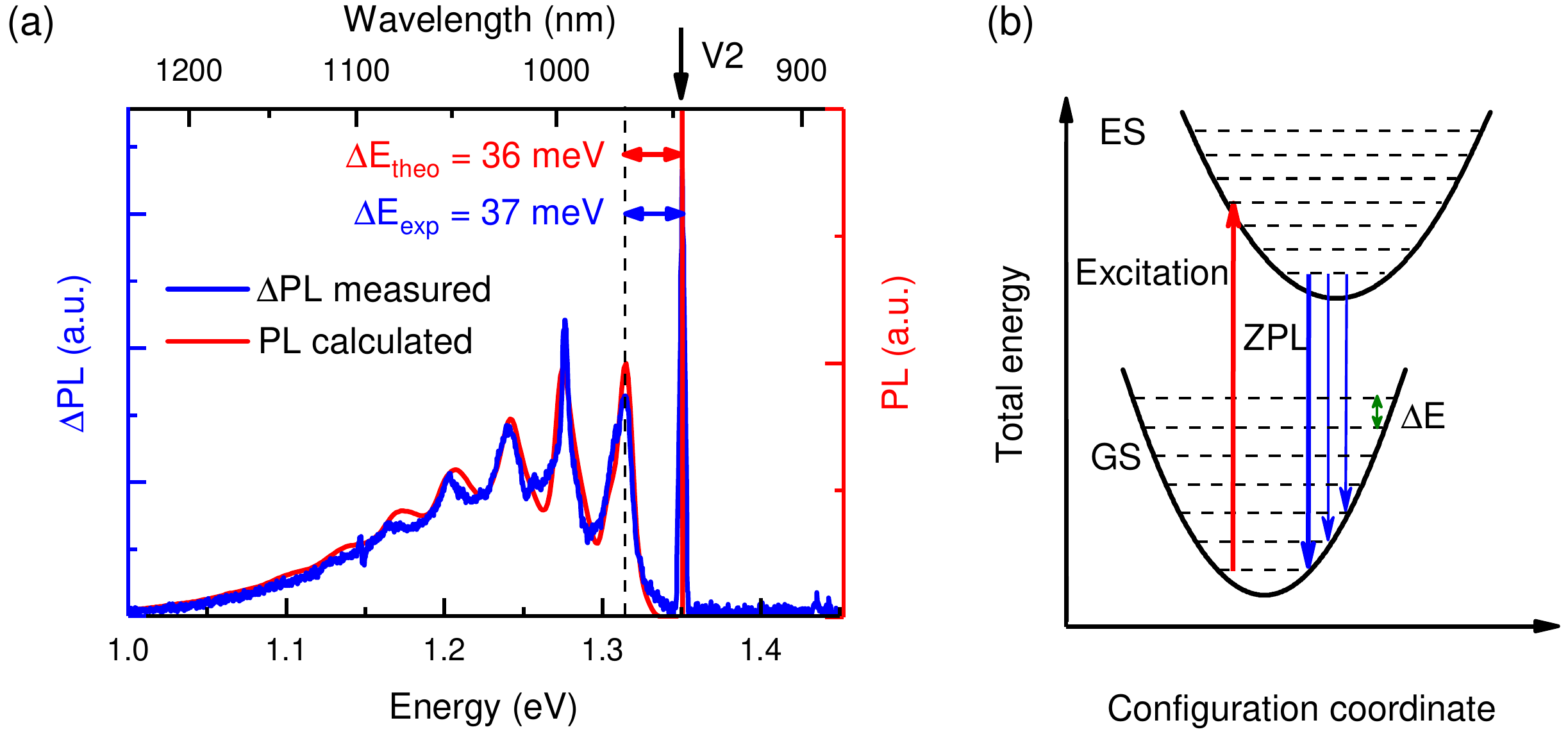}
\caption{(a) Low temperature ($T = 15 \, \mathrm{K}$)  $\mathrm{ \Delta PL}$  of V2 as a function of the detection wavelength at the MW frequency of 70 MHz.
The vertical axis for $\mathrm{PL}$ is the lock-in voltage of the modulated photodiode signal. The V2 ZPL (at $1.35 \, \mathrm{eV}$) and PSB  are clearly resolved.
The local vibrational energy read from two adjacent peaks is $\Delta E_{exp} = 37 \, \mathrm{meV}$. The red line represents the  calculation of the V2 PSB as described in the text.  (b) A configuration coordinate diagram for the V2 phonon modes. The blue arrows show transitions from the ES to different vibrational levels of the GS. } \label{fig4}
\end{figure*}

Next, we analyze how the change in the V2  $\mathrm{ \Delta PL}$ emission induced by the MW depends on the detection wavelength.  We set the MW frequency to $70 \, \mathrm{MHz}$ (Fig.~\ref{fig2}(c)) and the $\mathrm{ \Delta PL}$ spectrum at $T = 300 \, \mathrm{K}$ is presented in Fig.~\ref{fig3}(a). With  decreasing temperature, the $\mathrm{ \Delta PL}$ spectrum transfers to the ZPL and PSB consisting of seven equally-separated phonon replicas (Ph1 -- Ph7).  These spectra differ from the PL spectrum presented in Fig.~\ref{fig1}, which is composed of overlapping contributions from different defects. Especially, the V1 and V2 PSBs are merged together, making their separation different. In the  $\mathrm{ \Delta PL}$ measurements, the MW frequency of $70 \, \mathrm{MHz}$ is in the V2 spin resonance and, therefore, only the V2 PSB appears. 

The spectrally integrated V2 $\mathrm{ \Delta PL}$ as a function of temperature is presented in Fig.~\ref{fig3}(b).  The experimental data can be well reproduced using single activation energy \cite{Reshchikov:2011eo} 
\begin{equation}
 \mathrm{ \Delta PL} (T) = \frac{\mathrm{ \Delta PL}(0)}{1 + C \exp (-E_A / k_B T)} \,.
\label{Arrh}
\end{equation}
We obtain from the best fit (solid line in Fig.~\ref{fig3}(b)) the activation energy $E_A = 39 \pm 4 \, \mathrm{meV}$, which is equal within the experimental uncertainty to the local vibrational energy $\Delta E_{exp} = 37 \pm 4 \, \mathrm{meV}$, as discussed hereafter.  The unit-less coefficient $C = 9 \pm 2$ is determined by the ratio of different relaxation rates \cite{Reshchikov:2011eo}. The $\mathrm{ \Delta PL}$ is contributed by the spin polarization and the PL intensity. In order to separate these contributions, we plot in Fig.~\ref{fig3}(c) $\mathrm{ \Delta PL / PL}$ detected at the ZPL and the spectrally integrated PL. The experimental data for $\mathrm{ \Delta PL / PL}$ can be also well fitted to Eq.~(\ref{Arrh}) with the activation energy $E_A = 39 \pm 3 \, \mathrm{meV}$ and  $\mathrm{ \Delta PL} (0)$ replaced with  $\mathrm{ \Delta PL / PL} (0) = 0.7 \%$. This indicates that the integrated PL intensity of the V2 $\mathrm{V_{Si}}$ is nearly temperature independent up to $300 \, \mathrm{K}$. Indeed, this is in agreement with the integrated PL of Fig.~\ref{fig3}(c), where a small decrease with rising temperature can be attributed to the contribution of other defects with stronger temperature dependence. 

Figure~\ref{fig3}(a) clearly shows the PSB extends below $1.1 \, \mathrm{eV}$ (above $1150 \, \mathrm{nm}$). Thus, the  DW above 30\%  found in earlier experiments  \cite{Banks:2019je} is definitely overestimated.  Spectral integration of the experimental data in Fig.~\ref{fig3}(a) results in the DW factor of about 6\% for $T < 60 \, \mathrm{K}$. This value should be corrected by the spectrally dependent readout contrast. Given $\mathrm{ \Delta PL / PL}$ spectral dependence of Fig.~\ref{fig2}(d), the expected value for the DW factor falls between 6\% and 9\%. The DW decreases with temperature as shown in Fig.~\ref{fig3}(b).
As DW factor gives the fraction of elastic scattering, we might attribute the temperature reduction of its value to thermal motion effect \cite{Kittel2004} and multiphonon contributions.

We determine the local vibrational energy as the separation between two adjacent phonon peaks in $\Delta E_{exp} = 37 \, \mathrm{meV}$ as presented in  Fig.~\ref{fig4}(a). The PSB formation is schematically presented in  Fig.~\ref{fig4}(b). The radiative recombination between the ES and GS is accompanied by phonon emission. The energy dispersion of these phonons differs from the bulk phonon dispersion because of the broken translation symmetry in the vicinity of the $\mathrm{V_{Si}}$ defect. In the next section, we present detailed theoretical analysis of the local vibrational modes.

\section*{Theory}
\begin{figure}[ht]
\begin{center}
\includegraphics[width=8.5cm]{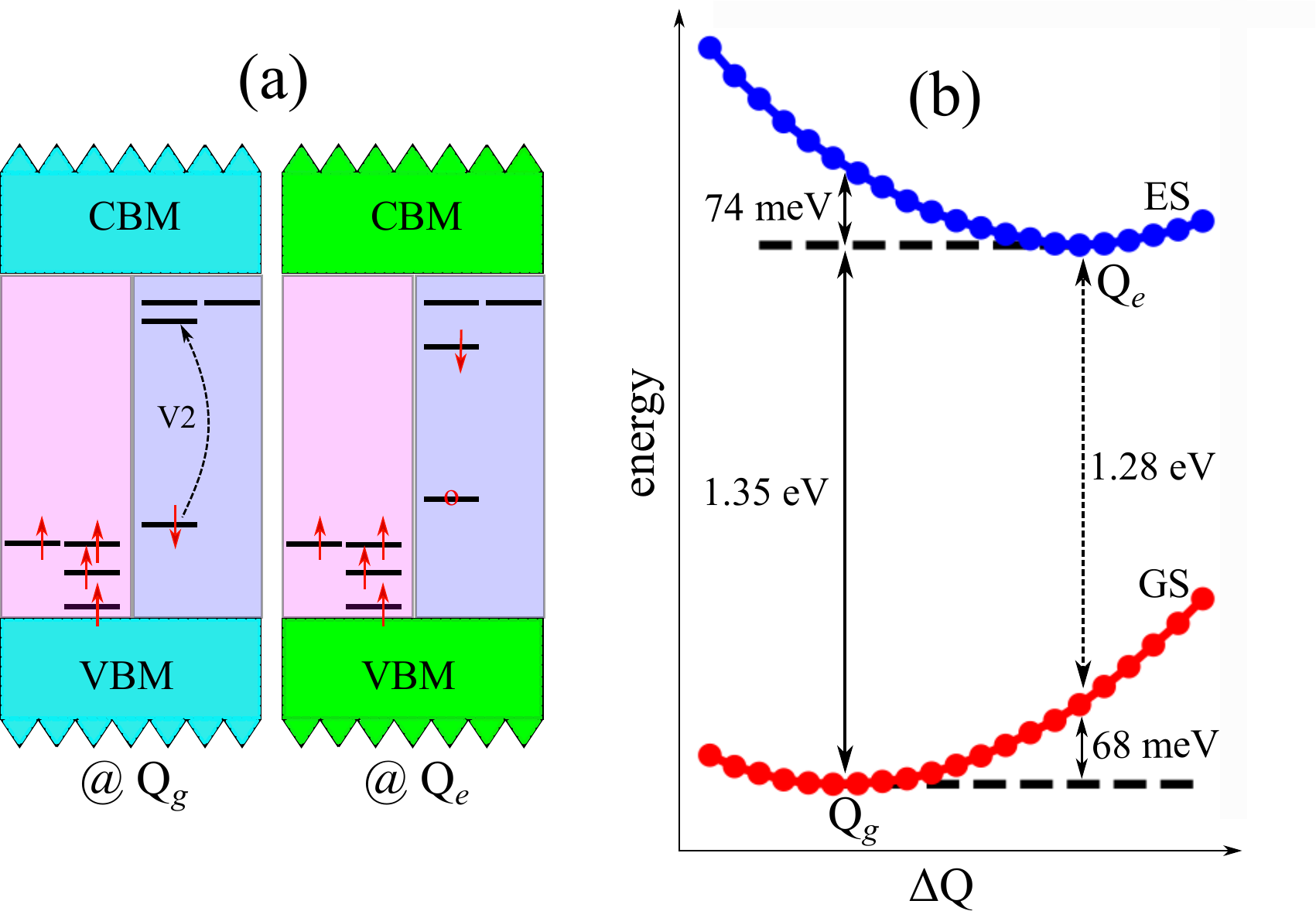}
\end{center}
\caption{(a) The arrangement of electronic states before (Q$_{g}$) and after (Q$_{e}$) excitations
calculated using HSE06 functional.
The states (dark lines) filled by electrons represented by arrows for spin-up and -down.
The hollow circle denotes hole.
(b) A schematic configuration coordinate diagram for the GS and ES showing energy scale for different transitions.}
\label{fig:ks}
\end{figure}
\begin{figure*}[t]
\centering
\includegraphics[width=17cm]{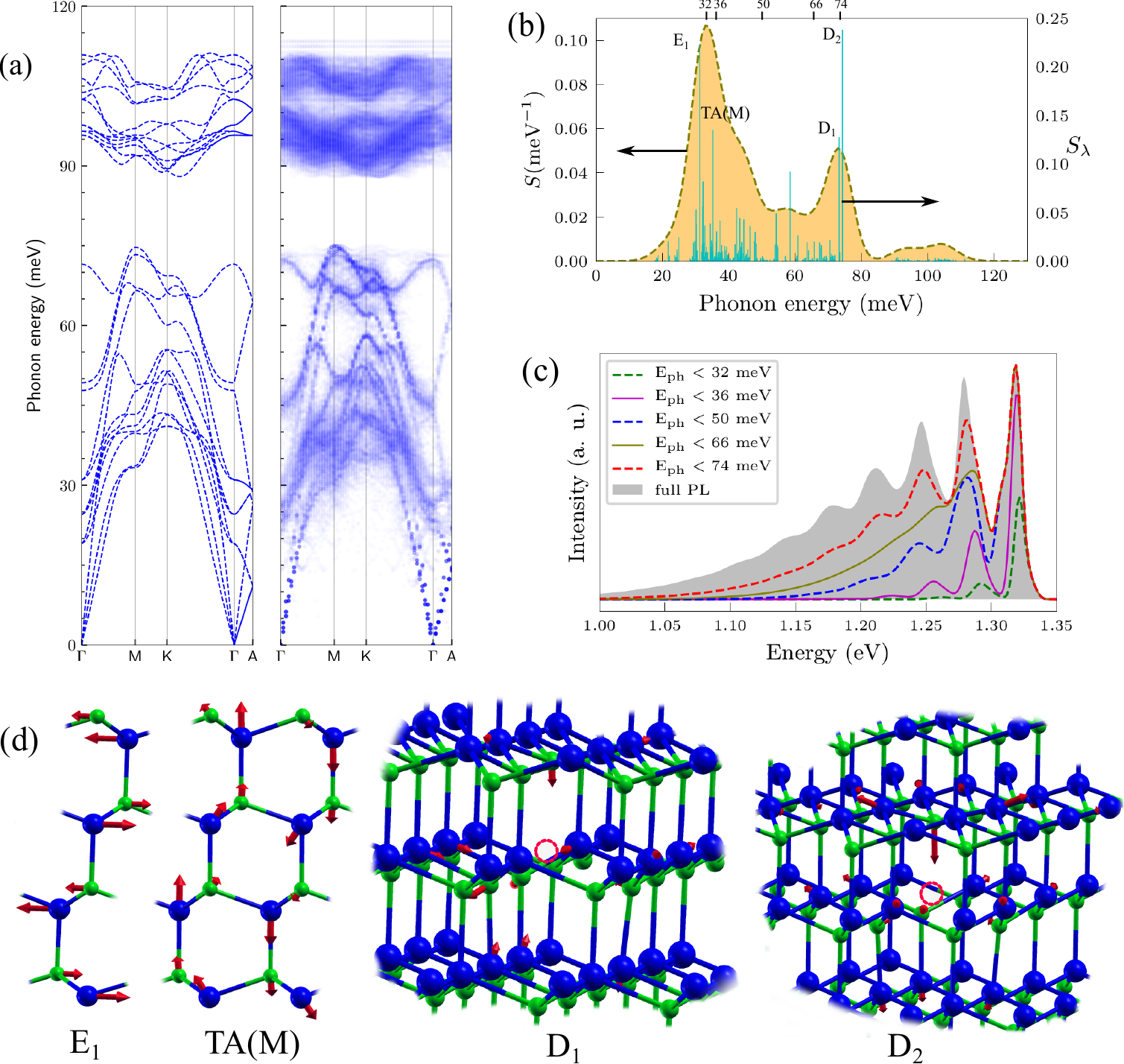}
\caption{(a) Pristine SiC phonon dispersion curves and the unfolded phonon curves of defective SiC along high symmetry directions.
(b) Electron-phonon spectral function accompanied by partial Huang-Rhys factors.
(c) The calculated partial PSB in the energy range of 1.0-1.35 eV.
(d) Schematic representations of atomic displacements of E$_{1}$, TA(M), D$_1$, and D$_{2}$ modes.
Blue and green balls denote silicon and carbon atoms, respectively.
Arrows are proportional to the displacements and come from the real part of the eigenvectors at the $\Gamma$-point.
The defect site is shown by a red circle.
}
\label{fig:defect}
\end{figure*}

We carried out DFT calculations as implemented in VASP \cite{VASP} code to determine defect properties,
configurational coordinate diagram, and vibrational modes which finally allow us to evaluate the PL lineshape.
A plane wave basis with a cutoff energy of 450 eV is employed to represent the electronic wave functions.
All structural relaxations and the vibrational properties were calculated using PBEsol \cite{PBEsol} exchange-correlation functional.
The geometry optimization continues until the energy differences and ionic forces are converged to less
than $10^{-6}$ eV and 0.01 eV/{\AA}, respectively.
The PL lineshape is calculated using the approach described in Ref.~\cite{Alkauskas:2014gq} and described in more detail below, which requires evaluation of the phonon spectra of the defective systems,
but obtaining converged spectra requires large supercells that are computationally very demanding.
Here, to speed up phonon calculations HIPHIVE \cite{HIPHIVE} package was used to extract interatomic force constants (IFCs). Second-order IFCs were constructed using the recursive feature elimination (RFE) optimizer by including pairs and triplets up to 4.2 and 3.6~{\AA}, respectively.
The modeled IFCs result in the validation root mean squared error of 13~meV/{\AA}.
The phonon frequencies and eigenvectors were finally assessed using PHONOPY software \cite{PHONOPY}.
To adjust the energy scales, i.e. the band gap and the position of the defect levels within,
we additionally used HSE06 \cite{hse06} to calculate total energies and Kohn-Sham levels.
HSE functional has been shown to reproduce intradefect transition energies very well \cite{Deak2010PRB}.
The calculated band gap of 3.25 eV is in excellent agreement to the experimental gap of 3.2 eV \cite{Lebedev1999}.

Theoretically, the determination of the average numbers of active phonons
during the optical transition for mode ${\lambda}$ with frequency $\omega_{\lambda}$
is given by the unitless partial Huang-Rhys (HR) factor $S_{\lambda}$ 
defined as  \cite{MARKHAM1959, Alkauskas:2014gq}
\begin{equation}
S_{\lambda} = \frac{1}{2 \hbar} \omega_{\lambda} {\Delta {Q}}_{\lambda}^{2},
\end{equation}
where
\begin{equation}\label{eq:DeltaQ}
\begin{aligned}
\Delta {Q}_{\lambda} = \sum_{\alpha} \sqrt{m_{\alpha}} [({\bf{R}}_{e, \alpha} - {\bf{R}}_{g, \alpha})\cdot {\bf{u}}_{\lambda}]. 
\end{aligned}
\end{equation}
Here, ${\bf{u}}_{\lambda}$ indicates the normalized displacement vector
corresponding to mode $\lambda$ and $m_{\alpha}$ is mass of atom $\alpha$.
${\bf{R}}_{g}$ and ${\bf{R}}_{e}$ are the atomic coordinates in the ground and excited state.
Thus, $\Delta {Q}_{\lambda}$ describes whether the vibrational mode
is parallel to the change in the atomic coordinates.
%
The fundamental spectral density of electron-phonon coupling can be determined as
\begin{equation}
S(\hbar \omega) \approx \sum_{\lambda} \frac{S_{\lambda}}{\sigma \sqrt{\pi}} e^{-\frac{(\hbar \omega - \hbar \omega_{\lambda})^{2}}{2 \sigma^{2}}},
\label{spectra_function}
\end{equation}
where a broadening parameter $\sigma = 5 \, \mathrm{meV}$ is considered.
It is worth mentioning that we assume the vibrational modes
in the ground state and the excited state to be identical,
and we use in Eq.\ \ref{eq:DeltaQ} the $\bf{u}$ evaluated in the ground state.
Once $S( \hbar \omega)$ is calculated, we make use of
the method of generating function \cite{Miyakawa1970, Alkauskas:2014gq} to derive the optical absorption spectrum
\begin{equation}
    L(\hbar \omega) = \frac{A \omega^3}{2 \pi} \int_{-\infty}^{+\infty} g(t) e^{i \omega t} dt,
\end{equation}
where the prefactor $A$ is the normalization constant and 
\begin{equation}
g(t) = e^{S(t) - S(0)} 
\end{equation}
is the generating function, where $S(t)$ is defined by
\begin{equation}
S(t) =\frac{1}{2} \int d(\hbar \omega) e^{-i \omega t} S(\hbar \omega).
\end{equation}

We modelled the $\mathrm{V_{Si}}$ defect in a large 400-atoms supercell using 2$\times$2$\times$2 meshes for $k$-point sampling.
V2 defect corresponds to $\mathrm{V_{Si}}$ at the h-site, in the $-1$ charge state, and
with spin $S=3/2$ \cite{Kraus:2013di}.
The HSE06-calculated Kohn-Sham levels are shown in Fig. \ref{fig:ks}(a), showing that
the V2 defect introduces several electronic states deep in the band gap.
The lowest energy electron configuration has the high spin state
with three unpaired electrons ($S=3/2$).
We model the excitation by moving spin-down electron from the highest occupied
to the first unoccupied electronic state, which is achieved
by fixing the occupations of the relevant states in the DFT calculations.
In the excited state,
the two states related to the excitation become closer while others remain the same.
From the DFT calculated total energies we can readily extract the configuration coordinate
diagram, as shown in Fig. \ref{fig:ks}(b).
We note that the potential energy curve is calculated using PBEsol, but the ZPL energy difference
($E(Q_g)-E(Q_e)$) is obtained using HSE06.
The emission energy of 1.28 eV and the ZPL of 1.35 eV
are in excellent agreement with the experimental values (the former corresponding
to PSB maximum).
Introduction of V2 defect also leads to expansion of the lattice by
$\Delta a = 0.083 \%$ and $\Delta c = 0.077 \%$.

The defects can induce new vibrational modes, that are either
resonant or anti-resonant with vibrational modes of the host crystal.
To map the vibrational modes on the same Brillouin zone as for the pristine SiC,
the unfolded phonon dispersion curve is illustrated in Fig. \ref{fig:defect}(a).
We find that
(i) the phonons between 20--40 meV and 90--110 meV are disturbed by the defects,
but still follow the dispersion of the bands seen in the pristine system, and
(ii) four localized vibrational modes (flat bands)  appear at energies
73.45, 74.4, 111.47, and 112.8 meV.
To find active phonons during emission process, the electron-phonon spectral function
is calculated as shown in Fig. \ref{fig:defect}(b).
Our calculation predicts that the PSB 
can be produced by a mix of about four phonon replicas: 
the double degenerate Raman-active E$_{1}$ mode with energy 31.3 meV,
the M-point transverse acoustic phonon active at 35.3 meV,
and two defect modes appearing at energies 73.45 and 74.4 meV, denoted as D$_1$ and D$_2$, respectively.
As illustrated in Fig \ref{fig:defect}(d),
in D$_1$ defect mode atoms up to second-nearest silicon neighbors from the vacancy center move,
while for D$_2$ case vibration is more intense and partially includes third nearest silicon atoms.

The partial HR factor ($S_{\lambda}$) is the average number of phonon $\lambda$ emitted during an optical transition.
We predict the total HR factor $S = 2.785$, which corresponds to the average number of phonons emitted
during an optical transition.
As a result, the weight of ZPL (DW factor) defined by $w_{ZPL} = \mathrm{e}^{-S}$
is 6.17$\%$, that is close to our measured value.
As was illustrated in Fig. \ref{fig4}(a), our prediction for PL lineshape is in full agreement with the experiments. The PSB with seven peaks falls off at around 1.1 eV.

Following the analysis of vibrational modes, both bulk and defect phonons should contribute to the PSB. To get more insight into lineshape, we calculate partial PSB lineshape.
To do this, we include phonons up to a specific energy in $S(\hbar \omega)$.
The phonon energy (E$_\mathrm{ph}$) is chosen based on the values of partial HR factors.
In this way, we can assess the contribution of different phonons into the total lineshape.
As seen in Fig. \ref{fig:defect}(c), the first peak shape is completed by adding phonons
up to 50 meV, but it quickly vanishes at lower energies. At this range of energy we have bulk phonons.
The position of the first peak is at 36 meV lower than ZPL that is in agreement with the experiments.
Clearly, the defect-induced phonons appeared around 74 meV are crucial
in shaping optical emission spectrum at low temperature.

\section*{Discussion and Conclusions}

Though the DW factor of 40\% and 30\% was reported for the V1 and V2 $\mathrm{V_{Si}}$ defects, respectively \cite{Nagy:2018ey, Banks:2019je}, the actual value is smaller due to the low detection efficiency at longer wavelength of the PSB. We experimentally estimated lower bound for the DW factor of 6\%--9\%.
On the other hand, the calculations also yielded DW factor of about 6\%, which suggests that the
real value is indeed in this range. Though this value is at least by a factor of 2 larger than that of the NV defect in diamond \cite{Alkauskas:2014gq}, thus coupling to an optical resonator is necessary to realize quantum repeaters. The vibrational energy of 37 meV and 36 meV from experiments and calculations, respectively, were also in close agreement. This is an important parameter, which determines at which temperature the phonon-assisted spin-lattice relaxation mechanism associated with local vibrational modes is activated \cite{Simin:2017iw}. 

To summarize, we have investigated the local vibrational structure of the V2 $\mathrm{V_{Si}}$ defect in a HPSI 4H-SiC wafer. The ODMR-based approach has enabled us to clearly separate the spectrally overlapped contribution from other intrinsic defects. We have found the resonant vibrational energy to be $36 \, \mathrm{meV}$ and estimated the lower bound for the DW factor to be 6\%. We have applied DFT-based methodology to calculate and analyze the PL lineshape. Beside that, we have established that the contribution into the optical emission process is narrowed down to dominant bulk and defect-induced phonons. All together, the perfect agreement between the experimental data and theoretical calculations shows that our approach can be applied to a large number of highly-promising optically addressable spin qubits in all stable SiC polytypes, including vacancies \cite{Riedel:2012jq}, divacancies  \cite{Falk:2013jq} and transition metal color centers \cite{Zargaleh:2016kx, Koehl:2017fd}. It is especially important when the spectral contribution from different defects is overlapped and cannot be separated otherwise. Exploring interaction of local vibrational modes with point defects, their spin, optical, mechanical and thermal properties can be predicted and understood, which is crucial by designing defect spins for quantum technologies.  We believe that our results have not only considerable interest for experimental research of defect-related spin-vibrational properties but also becomes an important tool to study a large variety of defects in wide-bandgap semiconductors and 2D materials \cite{Gottscholl:2019wd}. 

\section*{Acknowledgments}
This work has been supported by the German Research Foundation (DFG) under Grants  AS 310/5-1
and the Academy of Finland under Projects No.~286279 and 311058.
We also thank CSC-IT Center Science Ltd. Finland and PRACE (HLRS, Stuttgart, Germany)
for generous grants of computer time. The authors thank R.~Narkowicz and K.~Lenz for designing and characterization of the coplanar waveguides.



\begin{thebibliography}{66}%
\makeatletter
\providecommand \@ifxundefined [1]{%
 \@ifx{#1\undefined}
}%
\providecommand \@ifnum [1]{%
 \ifnum #1\expandafter \@firstoftwo
 \else \expandafter \@secondoftwo
 \fi
}%
\providecommand \@ifx [1]{%
 \ifx #1\expandafter \@firstoftwo
 \else \expandafter \@secondoftwo
 \fi
}%
\providecommand \natexlab [1]{#1}%
\providecommand \enquote  [1]{``#1''}%
\providecommand \bibnamefont  [1]{#1}%
\providecommand \bibfnamefont [1]{#1}%
\providecommand \citenamefont [1]{#1}%
\providecommand \href@noop [0]{\@secondoftwo}%
\providecommand \href [0]{\begingroup \@sanitize@url \@href}%
\providecommand \@href[1]{\@@startlink{#1}\@@href}%
\providecommand \@@href[1]{\endgroup#1\@@endlink}%
\providecommand \@sanitize@url [0]{\catcode `\\12\catcode `\$12\catcode
  `\&12\catcode `\#12\catcode `\^12\catcode `\_12\catcode `\%12\relax}%
\providecommand \@@startlink[1]{}%
\providecommand \@@endlink[0]{}%
\providecommand \url  [0]{\begingroup\@sanitize@url \@url }%
\providecommand \@url [1]{\endgroup\@href {#1}{\urlprefix }}%
\providecommand \urlprefix  [0]{URL }%
\providecommand \Eprint [0]{\href }%
\providecommand \doibase [0]{https://doi.org/}%
\providecommand \selectlanguage [0]{\@gobble}%
\providecommand \bibinfo  [0]{\@secondoftwo}%
\providecommand \bibfield  [0]{\@secondoftwo}%
\providecommand \translation [1]{[#1]}%
\providecommand \BibitemOpen [0]{}%
\providecommand \bibitemStop [0]{}%
\providecommand \bibitemNoStop [0]{.\EOS\space}%
\providecommand \EOS [0]{\spacefactor3000\relax}%
\providecommand \BibitemShut  [1]{\csname bibitem#1\endcsname}%
\let\auto@bib@innerbib\@empty
\bibitem [{\citenamefont {Baranov}\ \emph {et~al.}(2011)\citenamefont
  {Baranov}, \citenamefont {Bundakova}, \citenamefont {Soltamova},
  \citenamefont {Orlinskii}, \citenamefont {Borovykh}, \citenamefont
  {Zondervan}, \citenamefont {Verberk},\ and\ \citenamefont
  {Schmidt}}]{Baranov:2011ib}%
  \BibitemOpen
  \bibfield  {author} {\bibinfo {author} {\bibfnamefont {P.~G.}\ \bibnamefont
  {Baranov}}, \bibinfo {author} {\bibfnamefont {A.~P.}\ \bibnamefont
  {Bundakova}}, \bibinfo {author} {\bibfnamefont {A.~A.}\ \bibnamefont
  {Soltamova}}, \bibinfo {author} {\bibfnamefont {S.~B.}\ \bibnamefont
  {Orlinskii}}, \bibinfo {author} {\bibfnamefont {I.~V.}\ \bibnamefont
  {Borovykh}}, \bibinfo {author} {\bibfnamefont {R.}~\bibnamefont {Zondervan}},
  \bibinfo {author} {\bibfnamefont {R.}~\bibnamefont {Verberk}},\ and\ \bibinfo
  {author} {\bibfnamefont {J.}~\bibnamefont {Schmidt}},\ }\bibfield  {title}
  {\bibinfo {title} {{Silicon vacancy in SiC as a promising quantum system for
  single-defect and single-photon spectroscopy}},\ }\href@noop {} {\bibfield
  {journal} {\bibinfo  {journal} {Physical Review B}\ }\textbf {\bibinfo
  {volume} {83}},\ \bibinfo {pages} {125203} (\bibinfo {year}
  {2011})}\BibitemShut {NoStop}%
\bibitem [{\citenamefont {Koehl}\ \emph {et~al.}(2011)\citenamefont {Koehl},
  \citenamefont {Buckley}, \citenamefont {Heremans}, \citenamefont {Calusine},\
  and\ \citenamefont {Awschalom}}]{Koehl:2011fv}%
  \BibitemOpen
  \bibfield  {author} {\bibinfo {author} {\bibfnamefont {W.~F.}\ \bibnamefont
  {Koehl}}, \bibinfo {author} {\bibfnamefont {B.~B.}\ \bibnamefont {Buckley}},
  \bibinfo {author} {\bibfnamefont {F.~J.}\ \bibnamefont {Heremans}}, \bibinfo
  {author} {\bibfnamefont {G.}~\bibnamefont {Calusine}},\ and\ \bibinfo
  {author} {\bibfnamefont {D.~D.}\ \bibnamefont {Awschalom}},\ }\bibfield
  {title} {\bibinfo {title} {{Room temperature coherent control of defect spin
  qubits in silicon carbide}},\ }\href@noop {} {\bibfield  {journal} {\bibinfo
  {journal} {Nature}\ }\textbf {\bibinfo {volume} {479}},\ \bibinfo {pages}
  {84} (\bibinfo {year} {2011})}\BibitemShut {NoStop}%
\bibitem [{\citenamefont {Riedel}\ \emph {et~al.}(2012)\citenamefont {Riedel},
  \citenamefont {Fuchs}, \citenamefont {Kraus}, \citenamefont {V{\"a}th},
  \citenamefont {Sperlich}, \citenamefont {Dyakonov}, \citenamefont
  {Soltamova}, \citenamefont {Baranov}, \citenamefont {Ilyin},\ and\
  \citenamefont {Astakhov}}]{Riedel:2012jq}%
  \BibitemOpen
  \bibfield  {author} {\bibinfo {author} {\bibfnamefont {D.}~\bibnamefont
  {Riedel}}, \bibinfo {author} {\bibfnamefont {F.}~\bibnamefont {Fuchs}},
  \bibinfo {author} {\bibfnamefont {H.}~\bibnamefont {Kraus}}, \bibinfo
  {author} {\bibfnamefont {S.}~\bibnamefont {V{\"a}th}}, \bibinfo {author}
  {\bibfnamefont {A.}~\bibnamefont {Sperlich}}, \bibinfo {author}
  {\bibfnamefont {V.}~\bibnamefont {Dyakonov}}, \bibinfo {author}
  {\bibfnamefont {A.}~\bibnamefont {Soltamova}}, \bibinfo {author}
  {\bibfnamefont {P.}~\bibnamefont {Baranov}}, \bibinfo {author} {\bibfnamefont
  {V.}~\bibnamefont {Ilyin}},\ and\ \bibinfo {author} {\bibfnamefont {G.~V.}\
  \bibnamefont {Astakhov}},\ }\bibfield  {title} {\bibinfo {title} {{Resonant
  Addressing and Manipulation of Silicon Vacancy Qubits in Silicon Carbide}},\
  }\href@noop {} {\bibfield  {journal} {\bibinfo  {journal} {Physical Review
  Letters}\ }\textbf {\bibinfo {volume} {109}},\ \bibinfo {pages} {226402}
  (\bibinfo {year} {2012})}\BibitemShut {NoStop}%
\bibitem [{\citenamefont {Soltamov}\ \emph {et~al.}(2012)\citenamefont
  {Soltamov}, \citenamefont {Soltamova}, \citenamefont {Baranov},\ and\
  \citenamefont {Proskuryakov}}]{Soltamov:2012ey}%
  \BibitemOpen
  \bibfield  {author} {\bibinfo {author} {\bibfnamefont {V.~A.}\ \bibnamefont
  {Soltamov}}, \bibinfo {author} {\bibfnamefont {A.~A.}\ \bibnamefont
  {Soltamova}}, \bibinfo {author} {\bibfnamefont {P.~G.}\ \bibnamefont
  {Baranov}},\ and\ \bibinfo {author} {\bibfnamefont {I.~I.}\ \bibnamefont
  {Proskuryakov}},\ }\bibfield  {title} {\bibinfo {title} {{Room Temperature
  Coherent Spin Alignment of Silicon Vacancies in 4H- and 6H-SiC}},\
  }\href@noop {} {\bibfield  {journal} {\bibinfo  {journal} {Physical Review
  Letters}\ }\textbf {\bibinfo {volume} {108}},\ \bibinfo {pages} {226402}
  (\bibinfo {year} {2012})}\BibitemShut {NoStop}%
\bibitem [{\citenamefont {Kraus}\ \emph
  {et~al.}(2014{\natexlab{a}})\citenamefont {Kraus}, \citenamefont {Soltamov},
  \citenamefont {Riedel}, \citenamefont {V{\"a}th}, \citenamefont {Fuchs},
  \citenamefont {Sperlich}, \citenamefont {Baranov}, \citenamefont {Dyakonov},\
  and\ \citenamefont {Astakhov}}]{Kraus:2013di}%
  \BibitemOpen
  \bibfield  {author} {\bibinfo {author} {\bibfnamefont {H.}~\bibnamefont
  {Kraus}}, \bibinfo {author} {\bibfnamefont {V.~A.}\ \bibnamefont {Soltamov}},
  \bibinfo {author} {\bibfnamefont {D.}~\bibnamefont {Riedel}}, \bibinfo
  {author} {\bibfnamefont {S.}~\bibnamefont {V{\"a}th}}, \bibinfo {author}
  {\bibfnamefont {F.}~\bibnamefont {Fuchs}}, \bibinfo {author} {\bibfnamefont
  {A.}~\bibnamefont {Sperlich}}, \bibinfo {author} {\bibfnamefont {P.~G.}\
  \bibnamefont {Baranov}}, \bibinfo {author} {\bibfnamefont {V.}~\bibnamefont
  {Dyakonov}},\ and\ \bibinfo {author} {\bibfnamefont {G.~V.}\ \bibnamefont
  {Astakhov}},\ }\bibfield  {title} {\bibinfo {title} {{Room-temperature
  quantum microwave emitters based on spin defects in silicon carbide}},\
  }\href@noop {} {\bibfield  {journal} {\bibinfo  {journal} {Nature Physics}\
  }\textbf {\bibinfo {volume} {10}},\ \bibinfo {pages} {157} (\bibinfo {year}
  {2014}{\natexlab{a}})}\BibitemShut {NoStop}%
\bibitem [{\citenamefont {Castelletto}\ \emph {et~al.}(2013)\citenamefont
  {Castelletto}, \citenamefont {Johnson}, \citenamefont {Iv{\'a}dy},
  \citenamefont {Stavrias}, \citenamefont {Umeda}, \citenamefont {Gali},\ and\
  \citenamefont {Ohshima}}]{Castelletto:2013el}%
  \BibitemOpen
  \bibfield  {author} {\bibinfo {author} {\bibfnamefont {S.}~\bibnamefont
  {Castelletto}}, \bibinfo {author} {\bibfnamefont {B.~C.}\ \bibnamefont
  {Johnson}}, \bibinfo {author} {\bibfnamefont {V.}~\bibnamefont {Iv{\'a}dy}},
  \bibinfo {author} {\bibfnamefont {N.}~\bibnamefont {Stavrias}}, \bibinfo
  {author} {\bibfnamefont {T.}~\bibnamefont {Umeda}}, \bibinfo {author}
  {\bibfnamefont {A.}~\bibnamefont {Gali}},\ and\ \bibinfo {author}
  {\bibfnamefont {T.}~\bibnamefont {Ohshima}},\ }\bibfield  {title} {\bibinfo
  {title} {{A silicon carbide room-temperature single-photon source}},\
  }\href@noop {} {\bibfield  {journal} {\bibinfo  {journal} {Nature Materials}\
  }\textbf {\bibinfo {volume} {13}},\ \bibinfo {pages} {151} (\bibinfo {year}
  {2013})}\BibitemShut {NoStop}%
\bibitem [{\citenamefont {Christle}\ \emph {et~al.}(2015)\citenamefont
  {Christle}, \citenamefont {Falk}, \citenamefont {Andrich}, \citenamefont
  {Klimov}, \citenamefont {Hassan}, \citenamefont {Son}, \citenamefont
  {Janz{\'e}n}, \citenamefont {Ohshima},\ and\ \citenamefont
  {Awschalom}}]{Christle:2014ti}%
  \BibitemOpen
  \bibfield  {author} {\bibinfo {author} {\bibfnamefont {D.~J.}\ \bibnamefont
  {Christle}}, \bibinfo {author} {\bibfnamefont {A.~L.}\ \bibnamefont {Falk}},
  \bibinfo {author} {\bibfnamefont {P.}~\bibnamefont {Andrich}}, \bibinfo
  {author} {\bibfnamefont {P.~V.}\ \bibnamefont {Klimov}}, \bibinfo {author}
  {\bibfnamefont {J.~u.}\ \bibnamefont {Hassan}}, \bibinfo {author}
  {\bibfnamefont {N.~T.}\ \bibnamefont {Son}}, \bibinfo {author} {\bibfnamefont
  {E.}~\bibnamefont {Janz{\'e}n}}, \bibinfo {author} {\bibfnamefont
  {T.}~\bibnamefont {Ohshima}},\ and\ \bibinfo {author} {\bibfnamefont {D.~D.}\
  \bibnamefont {Awschalom}},\ }\bibfield  {title} {\bibinfo {title} {{Isolated
  electron spins in silicon carbide with millisecond coherence times}},\
  }\href@noop {} {\bibfield  {journal} {\bibinfo  {journal} {Nature Materials}\
  }\textbf {\bibinfo {volume} {14}},\ \bibinfo {pages} {160} (\bibinfo {year}
  {2015})}\BibitemShut {NoStop}%
\bibitem [{\citenamefont {Widmann}\ \emph {et~al.}(2015)\citenamefont
  {Widmann}, \citenamefont {Lee}, \citenamefont {Rendler}, \citenamefont {Son},
  \citenamefont {Fedder}, \citenamefont {Paik}, \citenamefont {Yang},
  \citenamefont {Zhao}, \citenamefont {Yang}, \citenamefont {Booker},
  \citenamefont {Denisenko}, \citenamefont {Jamali}, \citenamefont
  {Momenzadeh}, \citenamefont {Gerhardt}, \citenamefont {Ohshima},
  \citenamefont {Gali}, \citenamefont {Janz{\'e}n},\ and\ \citenamefont
  {Wrachtrup}}]{Widmann:2014ve}%
  \BibitemOpen
  \bibfield  {author} {\bibinfo {author} {\bibfnamefont {M.}~\bibnamefont
  {Widmann}}, \bibinfo {author} {\bibfnamefont {S.-Y.}\ \bibnamefont {Lee}},
  \bibinfo {author} {\bibfnamefont {T.}~\bibnamefont {Rendler}}, \bibinfo
  {author} {\bibfnamefont {N.~T.}\ \bibnamefont {Son}}, \bibinfo {author}
  {\bibfnamefont {H.}~\bibnamefont {Fedder}}, \bibinfo {author} {\bibfnamefont
  {S.}~\bibnamefont {Paik}}, \bibinfo {author} {\bibfnamefont {L.-P.}\
  \bibnamefont {Yang}}, \bibinfo {author} {\bibfnamefont {N.}~\bibnamefont
  {Zhao}}, \bibinfo {author} {\bibfnamefont {S.}~\bibnamefont {Yang}}, \bibinfo
  {author} {\bibfnamefont {I.}~\bibnamefont {Booker}}, \bibinfo {author}
  {\bibfnamefont {A.}~\bibnamefont {Denisenko}}, \bibinfo {author}
  {\bibfnamefont {M.}~\bibnamefont {Jamali}}, \bibinfo {author} {\bibfnamefont
  {S.~A.}\ \bibnamefont {Momenzadeh}}, \bibinfo {author} {\bibfnamefont
  {I.}~\bibnamefont {Gerhardt}}, \bibinfo {author} {\bibfnamefont
  {T.}~\bibnamefont {Ohshima}}, \bibinfo {author} {\bibfnamefont
  {A.}~\bibnamefont {Gali}}, \bibinfo {author} {\bibfnamefont {E.}~\bibnamefont
  {Janz{\'e}n}},\ and\ \bibinfo {author} {\bibfnamefont {J.}~\bibnamefont
  {Wrachtrup}},\ }\bibfield  {title} {\bibinfo {title} {{Coherent control of
  single spins in silicon carbide at room temperature}},\ }\href@noop {}
  {\bibfield  {journal} {\bibinfo  {journal} {Nature Materials}\ }\textbf
  {\bibinfo {volume} {14}},\ \bibinfo {pages} {164} (\bibinfo {year}
  {2015})}\BibitemShut {NoStop}%
\bibitem [{\citenamefont {Fuchs}\ \emph {et~al.}(2015)\citenamefont {Fuchs},
  \citenamefont {Stender}, \citenamefont {Trupke}, \citenamefont {Simin},
  \citenamefont {Pflaum}, \citenamefont {Dyakonov},\ and\ \citenamefont
  {Astakhov}}]{Fuchs:2015ii}%
  \BibitemOpen
  \bibfield  {author} {\bibinfo {author} {\bibfnamefont {F.}~\bibnamefont
  {Fuchs}}, \bibinfo {author} {\bibfnamefont {B.}~\bibnamefont {Stender}},
  \bibinfo {author} {\bibfnamefont {M.}~\bibnamefont {Trupke}}, \bibinfo
  {author} {\bibfnamefont {D.}~\bibnamefont {Simin}}, \bibinfo {author}
  {\bibfnamefont {J.}~\bibnamefont {Pflaum}}, \bibinfo {author} {\bibfnamefont
  {V.}~\bibnamefont {Dyakonov}},\ and\ \bibinfo {author} {\bibfnamefont
  {G.~V.}\ \bibnamefont {Astakhov}},\ }\bibfield  {title} {\bibinfo {title}
  {{Engineering near-infrared single-photon emitters with optically active
  spins in ultrapure silicon carbide}},\ }\href@noop {} {\bibfield  {journal}
  {\bibinfo  {journal} {Nature Communications}\ }\textbf {\bibinfo {volume}
  {6}},\ \bibinfo {pages} {7578} (\bibinfo {year} {2015})}\BibitemShut
  {NoStop}%
\bibitem [{\citenamefont {Kraus}\ \emph
  {et~al.}(2014{\natexlab{b}})\citenamefont {Kraus}, \citenamefont {Soltamov},
  \citenamefont {Fuchs}, \citenamefont {Simin}, \citenamefont {Sperlich},
  \citenamefont {Baranov}, \citenamefont {Astakhov},\ and\ \citenamefont
  {Dyakonov}}]{Kraus:2013vf}%
  \BibitemOpen
  \bibfield  {author} {\bibinfo {author} {\bibfnamefont {H.}~\bibnamefont
  {Kraus}}, \bibinfo {author} {\bibfnamefont {V.~A.}\ \bibnamefont {Soltamov}},
  \bibinfo {author} {\bibfnamefont {F.}~\bibnamefont {Fuchs}}, \bibinfo
  {author} {\bibfnamefont {D.}~\bibnamefont {Simin}}, \bibinfo {author}
  {\bibfnamefont {A.}~\bibnamefont {Sperlich}}, \bibinfo {author}
  {\bibfnamefont {P.~G.}\ \bibnamefont {Baranov}}, \bibinfo {author}
  {\bibfnamefont {G.~V.}\ \bibnamefont {Astakhov}},\ and\ \bibinfo {author}
  {\bibfnamefont {V.}~\bibnamefont {Dyakonov}},\ }\bibfield  {title} {\bibinfo
  {title} {{Magnetic field and temperature sensing with atomic-scale spin
  defects in silicon carbide}},\ }\href@noop {} {\bibfield  {journal} {\bibinfo
   {journal} {Scientific Reports}\ }\textbf {\bibinfo {volume} {4}},\ \bibinfo
  {pages} {5303} (\bibinfo {year} {2014}{\natexlab{b}})}\BibitemShut {NoStop}%
\bibitem [{\citenamefont {Simin}\ \emph {et~al.}(2015)\citenamefont {Simin},
  \citenamefont {Fuchs}, \citenamefont {Kraus}, \citenamefont {Sperlich},
  \citenamefont {Baranov}, \citenamefont {Astakhov},\ and\ \citenamefont
  {Dyakonov}}]{Simin:2015dn}%
  \BibitemOpen
  \bibfield  {author} {\bibinfo {author} {\bibfnamefont {D.}~\bibnamefont
  {Simin}}, \bibinfo {author} {\bibfnamefont {F.}~\bibnamefont {Fuchs}},
  \bibinfo {author} {\bibfnamefont {H.}~\bibnamefont {Kraus}}, \bibinfo
  {author} {\bibfnamefont {A.}~\bibnamefont {Sperlich}}, \bibinfo {author}
  {\bibfnamefont {P.~G.}\ \bibnamefont {Baranov}}, \bibinfo {author}
  {\bibfnamefont {G.~V.}\ \bibnamefont {Astakhov}},\ and\ \bibinfo {author}
  {\bibfnamefont {V.}~\bibnamefont {Dyakonov}},\ }\bibfield  {title} {\bibinfo
  {title} {{High-Precision Angle-Resolved Magnetometry with Uniaxial Quantum
  Centers in Silicon Carbide}},\ }\href@noop {} {\bibfield  {journal} {\bibinfo
   {journal} {Physical Review Applied}\ }\textbf {\bibinfo {volume} {4}},\
  \bibinfo {pages} {014009} (\bibinfo {year} {2015})}\BibitemShut {NoStop}%
\bibitem [{\citenamefont {Simin}\ \emph {et~al.}(2016)\citenamefont {Simin},
  \citenamefont {Soltamov}, \citenamefont {Poshakinskiy}, \citenamefont
  {Anisimov}, \citenamefont {Babunts}, \citenamefont {Tolmachev}, \citenamefont
  {Mokhov}, \citenamefont {Trupke}, \citenamefont {Tarasenko}, \citenamefont
  {Sperlich}, \citenamefont {Baranov}, \citenamefont {Dyakonov},\ and\
  \citenamefont {Astakhov}}]{Simin:2016cp}%
  \BibitemOpen
  \bibfield  {author} {\bibinfo {author} {\bibfnamefont {D.}~\bibnamefont
  {Simin}}, \bibinfo {author} {\bibfnamefont {V.~A.}\ \bibnamefont {Soltamov}},
  \bibinfo {author} {\bibfnamefont {A.~V.}\ \bibnamefont {Poshakinskiy}},
  \bibinfo {author} {\bibfnamefont {A.~N.}\ \bibnamefont {Anisimov}}, \bibinfo
  {author} {\bibfnamefont {R.~A.}\ \bibnamefont {Babunts}}, \bibinfo {author}
  {\bibfnamefont {D.~O.}\ \bibnamefont {Tolmachev}}, \bibinfo {author}
  {\bibfnamefont {E.~N.}\ \bibnamefont {Mokhov}}, \bibinfo {author}
  {\bibfnamefont {M.}~\bibnamefont {Trupke}}, \bibinfo {author} {\bibfnamefont
  {S.~A.}\ \bibnamefont {Tarasenko}}, \bibinfo {author} {\bibfnamefont
  {A.}~\bibnamefont {Sperlich}}, \bibinfo {author} {\bibfnamefont {P.~G.}\
  \bibnamefont {Baranov}}, \bibinfo {author} {\bibfnamefont {V.}~\bibnamefont
  {Dyakonov}},\ and\ \bibinfo {author} {\bibfnamefont {G.~V.}\ \bibnamefont
  {Astakhov}},\ }\bibfield  {title} {\bibinfo {title} {{All-Optical dc
  Nanotesla Magnetometry Using Silicon Vacancy Fine Structure in Isotopically
  Purified Silicon Carbide}},\ }\href@noop {} {\bibfield  {journal} {\bibinfo
  {journal} {Physical Review X}\ }\textbf {\bibinfo {volume} {6}},\ \bibinfo
  {pages} {031014} (\bibinfo {year} {2016})}\BibitemShut {NoStop}%
\bibitem [{\citenamefont {Niethammer}\ \emph {et~al.}(2016)\citenamefont
  {Niethammer}, \citenamefont {Widmann}, \citenamefont {Lee}, \citenamefont
  {Stenberg}, \citenamefont {Kordina}, \citenamefont {Ohshima}, \citenamefont
  {Son}, \citenamefont {Janz{\'e}n},\ and\ \citenamefont
  {Wrachtrup}}]{Niethammer:2016bc}%
  \BibitemOpen
  \bibfield  {author} {\bibinfo {author} {\bibfnamefont {M.}~\bibnamefont
  {Niethammer}}, \bibinfo {author} {\bibfnamefont {M.}~\bibnamefont {Widmann}},
  \bibinfo {author} {\bibfnamefont {S.-Y.}\ \bibnamefont {Lee}}, \bibinfo
  {author} {\bibfnamefont {P.}~\bibnamefont {Stenberg}}, \bibinfo {author}
  {\bibfnamefont {O.}~\bibnamefont {Kordina}}, \bibinfo {author} {\bibfnamefont
  {T.}~\bibnamefont {Ohshima}}, \bibinfo {author} {\bibfnamefont {N.~T.}\
  \bibnamefont {Son}}, \bibinfo {author} {\bibfnamefont {E.}~\bibnamefont
  {Janz{\'e}n}},\ and\ \bibinfo {author} {\bibfnamefont {J.}~\bibnamefont
  {Wrachtrup}},\ }\bibfield  {title} {\bibinfo {title} {{Vector Magnetometry
  Using Silicon Vacancies in 4H-SiC Under Ambient Conditions}},\ }\href@noop {}
  {\bibfield  {journal} {\bibinfo  {journal} {Physical Review Applied}\
  }\textbf {\bibinfo {volume} {6}},\ \bibinfo {pages} {034001} (\bibinfo {year}
  {2016})}\BibitemShut {NoStop}%
\bibitem [{\citenamefont {Cochrane}\ \emph {et~al.}(2016)\citenamefont
  {Cochrane}, \citenamefont {Blacksberg}, \citenamefont {Anders},\ and\
  \citenamefont {Lenahan}}]{Cochrane:2016dd}%
  \BibitemOpen
  \bibfield  {author} {\bibinfo {author} {\bibfnamefont {C.~J.}\ \bibnamefont
  {Cochrane}}, \bibinfo {author} {\bibfnamefont {J.}~\bibnamefont
  {Blacksberg}}, \bibinfo {author} {\bibfnamefont {M.~A.}\ \bibnamefont
  {Anders}},\ and\ \bibinfo {author} {\bibfnamefont {P.~M.}\ \bibnamefont
  {Lenahan}},\ }\bibfield  {title} {\bibinfo {title} {{Vectorized magnetometer
  for space applications using electrical readout of atomic scale defects in
  silicon carbide}},\ }\href@noop {} {\bibfield  {journal} {\bibinfo  {journal}
  {Scientific Reports}\ }\textbf {\bibinfo {volume} {6}},\ \bibinfo {pages}
  {37077} (\bibinfo {year} {2016})}\BibitemShut {NoStop}%
\bibitem [{\citenamefont {Soykal}\ and\ \citenamefont
  {Reinecke}(2017)}]{Soykal:2016tk}%
  \BibitemOpen
  \bibfield  {author} {\bibinfo {author} {\bibfnamefont {{\"O}.~O.}\
  \bibnamefont {Soykal}}\ and\ \bibinfo {author} {\bibfnamefont {T.~L.}\
  \bibnamefont {Reinecke}},\ }\bibfield  {title} {\bibinfo {title} {{Quantum
  metrology with a single spin-3/2 defect in silicon carbide}},\ }\href@noop {}
  {\bibfield  {journal} {\bibinfo  {journal} {Physical Review B}\ }\textbf
  {\bibinfo {volume} {95}},\ \bibinfo {pages} {081405} (\bibinfo {year}
  {2017})}\BibitemShut {NoStop}%
\bibitem [{\citenamefont {Soltamov}\ \emph {et~al.}(2019)\citenamefont
  {Soltamov}, \citenamefont {Kasper}, \citenamefont {Poshakinskiy},
  \citenamefont {Anisimov}, \citenamefont {Mokhov}, \citenamefont {Sperlich},
  \citenamefont {Tarasenko}, \citenamefont {Baranov}, \citenamefont
  {Astakhov},\ and\ \citenamefont {Dyakonov}}]{Soltamov:2019hr}%
  \BibitemOpen
  \bibfield  {author} {\bibinfo {author} {\bibfnamefont {V.~A.}\ \bibnamefont
  {Soltamov}}, \bibinfo {author} {\bibfnamefont {C.}~\bibnamefont {Kasper}},
  \bibinfo {author} {\bibfnamefont {A.~V.}\ \bibnamefont {Poshakinskiy}},
  \bibinfo {author} {\bibfnamefont {A.~N.}\ \bibnamefont {Anisimov}}, \bibinfo
  {author} {\bibfnamefont {E.~N.}\ \bibnamefont {Mokhov}}, \bibinfo {author}
  {\bibfnamefont {A.}~\bibnamefont {Sperlich}}, \bibinfo {author}
  {\bibfnamefont {S.~A.}\ \bibnamefont {Tarasenko}}, \bibinfo {author}
  {\bibfnamefont {P.~G.}\ \bibnamefont {Baranov}}, \bibinfo {author}
  {\bibfnamefont {G.~V.}\ \bibnamefont {Astakhov}},\ and\ \bibinfo {author}
  {\bibfnamefont {V.}~\bibnamefont {Dyakonov}},\ }\bibfield  {title} {\bibinfo
  {title} {{Excitation and coherent control of spin qudit modes in silicon
  carbide at room temperature}},\ }\href@noop {} {\bibfield  {journal}
  {\bibinfo  {journal} {Nature Communications}\ }\textbf {\bibinfo {volume}
  {10}},\ \bibinfo {pages} {1678} (\bibinfo {year} {2019})}\BibitemShut
  {NoStop}%
\bibitem [{\citenamefont {Wolfowicz}\ \emph {et~al.}(2019)\citenamefont
  {Wolfowicz}, \citenamefont {Anderson}, \citenamefont {Whiteley},\ and\
  \citenamefont {Awschalom}}]{Wolfowicz:2019cz}%
  \BibitemOpen
  \bibfield  {author} {\bibinfo {author} {\bibfnamefont {G.}~\bibnamefont
  {Wolfowicz}}, \bibinfo {author} {\bibfnamefont {C.~P.}\ \bibnamefont
  {Anderson}}, \bibinfo {author} {\bibfnamefont {S.~J.}\ \bibnamefont
  {Whiteley}},\ and\ \bibinfo {author} {\bibfnamefont {D.~D.}\ \bibnamefont
  {Awschalom}},\ }\bibfield  {title} {\bibinfo {title} {{Heterodyne detection
  of radio-frequency electric fields using point defects in silicon carbide}},\
  }\href@noop {} {\bibfield  {journal} {\bibinfo  {journal} {Applied Physics
  Letters}\ }\textbf {\bibinfo {volume} {115}},\ \bibinfo {pages} {043105}
  (\bibinfo {year} {2019})}\BibitemShut {NoStop}%
\bibitem [{\citenamefont {Anisimov}\ \emph {et~al.}(2016)\citenamefont
  {Anisimov}, \citenamefont {Simin}, \citenamefont {Soltamov}, \citenamefont
  {Lebedev}, \citenamefont {Baranov}, \citenamefont {Astakhov},\ and\
  \citenamefont {Dyakonov}}]{Anisimov:2016er}%
  \BibitemOpen
  \bibfield  {author} {\bibinfo {author} {\bibfnamefont {A.~N.}\ \bibnamefont
  {Anisimov}}, \bibinfo {author} {\bibfnamefont {D.}~\bibnamefont {Simin}},
  \bibinfo {author} {\bibfnamefont {V.~A.}\ \bibnamefont {Soltamov}}, \bibinfo
  {author} {\bibfnamefont {S.~P.}\ \bibnamefont {Lebedev}}, \bibinfo {author}
  {\bibfnamefont {P.~G.}\ \bibnamefont {Baranov}}, \bibinfo {author}
  {\bibfnamefont {G.~V.}\ \bibnamefont {Astakhov}},\ and\ \bibinfo {author}
  {\bibfnamefont {V.}~\bibnamefont {Dyakonov}},\ }\bibfield  {title} {\bibinfo
  {title} {{Optical thermometry based on level anticrossing in silicon
  carbide}},\ }\href@noop {} {\bibfield  {journal} {\bibinfo  {journal}
  {Scientific Reports}\ }\textbf {\bibinfo {volume} {6}},\ \bibinfo {pages}
  {33301} (\bibinfo {year} {2016})}\BibitemShut {NoStop}%
\bibitem [{\citenamefont {Zhou}\ \emph {et~al.}(2017)\citenamefont {Zhou},
  \citenamefont {Wang}, \citenamefont {Zhang}, \citenamefont {Li},
  \citenamefont {Cai},\ and\ \citenamefont {Gao}}]{Zhou:2017cca}%
  \BibitemOpen
  \bibfield  {author} {\bibinfo {author} {\bibfnamefont {Y.}~\bibnamefont
  {Zhou}}, \bibinfo {author} {\bibfnamefont {J.}~\bibnamefont {Wang}}, \bibinfo
  {author} {\bibfnamefont {X.}~\bibnamefont {Zhang}}, \bibinfo {author}
  {\bibfnamefont {K.}~\bibnamefont {Li}}, \bibinfo {author} {\bibfnamefont
  {J.}~\bibnamefont {Cai}},\ and\ \bibinfo {author} {\bibfnamefont
  {W.}~\bibnamefont {Gao}},\ }\bibfield  {title} {\bibinfo {title}
  {{Self-Protected Thermometry with Infrared Photons and Defect Spins in
  Silicon Carbide}},\ }\href@noop {} {\bibfield  {journal} {\bibinfo  {journal}
  {Physical Review Applied}\ }\textbf {\bibinfo {volume} {8}},\ \bibinfo
  {pages} {044015} (\bibinfo {year} {2017})}\BibitemShut {NoStop}%
\bibitem [{\citenamefont {Yang}\ \emph {et~al.}(2014)\citenamefont {Yang},
  \citenamefont {Burk}, \citenamefont {Widmann}, \citenamefont {Lee},
  \citenamefont {Wrachtrup},\ and\ \citenamefont {Zhao}}]{Yang:2014kqa}%
  \BibitemOpen
  \bibfield  {author} {\bibinfo {author} {\bibfnamefont {L.-P.}\ \bibnamefont
  {Yang}}, \bibinfo {author} {\bibfnamefont {C.}~\bibnamefont {Burk}}, \bibinfo
  {author} {\bibfnamefont {M.}~\bibnamefont {Widmann}}, \bibinfo {author}
  {\bibfnamefont {S.-Y.}\ \bibnamefont {Lee}}, \bibinfo {author} {\bibfnamefont
  {J.}~\bibnamefont {Wrachtrup}},\ and\ \bibinfo {author} {\bibfnamefont
  {N.}~\bibnamefont {Zhao}},\ }\bibfield  {title} {\bibinfo {title} {{Electron
  spin decoherence in silicon carbide nuclear spin bath}},\ }\href@noop {}
  {\bibfield  {journal} {\bibinfo  {journal} {Physical Review B}\ }\textbf
  {\bibinfo {volume} {90}},\ \bibinfo {pages} {241203} (\bibinfo {year}
  {2014})}\BibitemShut {NoStop}%
\bibitem [{\citenamefont {Carter}\ \emph {et~al.}(2015)\citenamefont {Carter},
  \citenamefont {Soykal}, \citenamefont {Dev}, \citenamefont {Economou},\ and\
  \citenamefont {Glaser}}]{Carter:2015vc}%
  \BibitemOpen
  \bibfield  {author} {\bibinfo {author} {\bibfnamefont {S.~G.}\ \bibnamefont
  {Carter}}, \bibinfo {author} {\bibfnamefont {{\"O}.~O.}\ \bibnamefont
  {Soykal}}, \bibinfo {author} {\bibfnamefont {P.}~\bibnamefont {Dev}},
  \bibinfo {author} {\bibfnamefont {S.~E.}\ \bibnamefont {Economou}},\ and\
  \bibinfo {author} {\bibfnamefont {E.~R.}\ \bibnamefont {Glaser}},\ }\bibfield
   {title} {\bibinfo {title} {{Spin coherence and echo modulation of the
  silicon vacancy in 4H-SiC at room temperature}},\ }\href@noop {} {\bibfield
  {journal} {\bibinfo  {journal} {Physical Review B}\ }\textbf {\bibinfo
  {volume} {92}},\ \bibinfo {pages} {161202} (\bibinfo {year}
  {2015})}\BibitemShut {NoStop}%
\bibitem [{\citenamefont {Seo}\ \emph {et~al.}(2016)\citenamefont {Seo},
  \citenamefont {Falk}, \citenamefont {Klimov}, \citenamefont {Miao},
  \citenamefont {Galli},\ and\ \citenamefont {Awschalom}}]{Seo:2016ey}%
  \BibitemOpen
  \bibfield  {author} {\bibinfo {author} {\bibfnamefont {H.}~\bibnamefont
  {Seo}}, \bibinfo {author} {\bibfnamefont {A.~L.}\ \bibnamefont {Falk}},
  \bibinfo {author} {\bibfnamefont {P.~V.}\ \bibnamefont {Klimov}}, \bibinfo
  {author} {\bibfnamefont {K.~C.}\ \bibnamefont {Miao}}, \bibinfo {author}
  {\bibfnamefont {G.}~\bibnamefont {Galli}},\ and\ \bibinfo {author}
  {\bibfnamefont {D.~D.}\ \bibnamefont {Awschalom}},\ }\bibfield  {title}
  {\bibinfo {title} {{Quantum decoherence dynamics of divacancy spins in
  silicon carbide}},\ }\href@noop {} {\bibfield  {journal} {\bibinfo  {journal}
  {Nature Communications}\ }\textbf {\bibinfo {volume} {7}},\ \bibinfo {pages}
  {12935} (\bibinfo {year} {2016})}\BibitemShut {NoStop}%
\bibitem [{\citenamefont {Embley}\ \emph {et~al.}(2017)\citenamefont {Embley},
  \citenamefont {Colton}, \citenamefont {Miller}, \citenamefont {Morris},
  \citenamefont {Meehan}, \citenamefont {Crossen}, \citenamefont {Weaver},
  \citenamefont {Glaser},\ and\ \citenamefont {Carter}}]{Embley:2017bf}%
  \BibitemOpen
  \bibfield  {author} {\bibinfo {author} {\bibfnamefont {J.~S.}\ \bibnamefont
  {Embley}}, \bibinfo {author} {\bibfnamefont {J.~S.}\ \bibnamefont {Colton}},
  \bibinfo {author} {\bibfnamefont {K.~G.}\ \bibnamefont {Miller}}, \bibinfo
  {author} {\bibfnamefont {M.~A.}\ \bibnamefont {Morris}}, \bibinfo {author}
  {\bibfnamefont {M.}~\bibnamefont {Meehan}}, \bibinfo {author} {\bibfnamefont
  {S.~L.}\ \bibnamefont {Crossen}}, \bibinfo {author} {\bibfnamefont {B.~D.}\
  \bibnamefont {Weaver}}, \bibinfo {author} {\bibfnamefont {E.~R.}\
  \bibnamefont {Glaser}},\ and\ \bibinfo {author} {\bibfnamefont {S.~G.}\
  \bibnamefont {Carter}},\ }\bibfield  {title} {\bibinfo {title} {{Electron
  spin coherence of silicon vacancies in proton-irradiated 4H-SiC}},\
  }\href@noop {} {\bibfield  {journal} {\bibinfo  {journal} {Physical Review
  B}\ }\textbf {\bibinfo {volume} {95}},\ \bibinfo {pages} {045206} (\bibinfo
  {year} {2017})}\BibitemShut {NoStop}%
\bibitem [{\citenamefont {Simin}\ \emph {et~al.}(2017)\citenamefont {Simin},
  \citenamefont {Kraus}, \citenamefont {Sperlich}, \citenamefont {Ohshima},
  \citenamefont {Astakhov},\ and\ \citenamefont {Dyakonov}}]{Simin:2017iw}%
  \BibitemOpen
  \bibfield  {author} {\bibinfo {author} {\bibfnamefont {D.}~\bibnamefont
  {Simin}}, \bibinfo {author} {\bibfnamefont {H.}~\bibnamefont {Kraus}},
  \bibinfo {author} {\bibfnamefont {A.}~\bibnamefont {Sperlich}}, \bibinfo
  {author} {\bibfnamefont {T.}~\bibnamefont {Ohshima}}, \bibinfo {author}
  {\bibfnamefont {G.~V.}\ \bibnamefont {Astakhov}},\ and\ \bibinfo {author}
  {\bibfnamefont {V.}~\bibnamefont {Dyakonov}},\ }\bibfield  {title} {\bibinfo
  {title} {{Locking of electron spin coherence above 20 ms in natural silicon
  carbide}},\ }\href@noop {} {\bibfield  {journal} {\bibinfo  {journal}
  {Physical Review B}\ }\textbf {\bibinfo {volume} {95}},\ \bibinfo {pages}
  {161201(R)} (\bibinfo {year} {2017})}\BibitemShut {NoStop}%
\bibitem [{\citenamefont {Fischer}\ \emph {et~al.}(2018)\citenamefont
  {Fischer}, \citenamefont {Sperlich}, \citenamefont {Kraus}, \citenamefont
  {Ohshima}, \citenamefont {Astakhov},\ and\ \citenamefont
  {Dyakonov}}]{Fischer:2018fj}%
  \BibitemOpen
  \bibfield  {author} {\bibinfo {author} {\bibfnamefont {M.}~\bibnamefont
  {Fischer}}, \bibinfo {author} {\bibfnamefont {A.}~\bibnamefont {Sperlich}},
  \bibinfo {author} {\bibfnamefont {H.}~\bibnamefont {Kraus}}, \bibinfo
  {author} {\bibfnamefont {T.}~\bibnamefont {Ohshima}}, \bibinfo {author}
  {\bibfnamefont {G.~V.}\ \bibnamefont {Astakhov}},\ and\ \bibinfo {author}
  {\bibfnamefont {V.}~\bibnamefont {Dyakonov}},\ }\bibfield  {title} {\bibinfo
  {title} {{Highly Efficient Optical Pumping of Spin Defects in Silicon Carbide
  for Stimulated Microwave Emission}},\ }\href@noop {} {\bibfield  {journal}
  {\bibinfo  {journal} {Physical Review Applied}\ }\textbf {\bibinfo {volume}
  {9}},\ \bibinfo {pages} {2126} (\bibinfo {year} {2018})}\BibitemShut
  {NoStop}%
\bibitem [{\citenamefont {Economou}\ and\ \citenamefont
  {Dev}(2016)}]{Economou:2016bp}%
  \BibitemOpen
  \bibfield  {author} {\bibinfo {author} {\bibfnamefont {S.~E.}\ \bibnamefont
  {Economou}}\ and\ \bibinfo {author} {\bibfnamefont {P.}~\bibnamefont {Dev}},\
  }\bibfield  {title} {\bibinfo {title} {{Spin-photon entanglement interfaces
  in silicon carbide defect centers}},\ }\href@noop {} {\bibfield  {journal}
  {\bibinfo  {journal} {Nanotechnology}\ }\textbf {\bibinfo {volume} {27}},\
  \bibinfo {pages} {504001} (\bibinfo {year} {2016})}\BibitemShut {NoStop}%
\bibitem [{\citenamefont {Christle}\ \emph {et~al.}(2017)\citenamefont
  {Christle}, \citenamefont {Klimov}, \citenamefont {de~las Casas},
  \citenamefont {Sz{\'a}sz}, \citenamefont {Iv{\'a}dy}, \citenamefont {Jokubavicius},
  \citenamefont {Ul~Hassan}, \citenamefont {Syv{\"a}j{\"a}rvi}, \citenamefont {Koehl},
  \citenamefont {Ohshima}, \citenamefont {Son}, \citenamefont {Janz{\'e}n},
  \citenamefont {Gali},\ and\ \citenamefont {Awschalom}}]{Christle:2017tq}%
  \BibitemOpen
  \bibfield  {author} {\bibinfo {author} {\bibfnamefont {D.~J.}\ \bibnamefont
  {Christle}}, \bibinfo {author} {\bibfnamefont {P.~V.}\ \bibnamefont
  {Klimov}}, \bibinfo {author} {\bibfnamefont {C.~F.}\ \bibnamefont {de~las
  Casas}}, \bibinfo {author} {\bibfnamefont {K.}\ \bibnamefont {Sz{\'a}sz}},
  \bibinfo {author} {\bibfnamefont {V.}~\bibnamefont {Iv{\'a}dy}}, \bibinfo
  {author} {\bibfnamefont {V.}~\bibnamefont {Jokubavicius}}, \bibinfo {author}
  {\bibfnamefont {J.}~\bibnamefont {Ul~Hassan}}, \bibinfo {author}
  {\bibfnamefont {M.}~\bibnamefont {Syv{\"a}j{\"a}rvi}}, \bibinfo {author}
  {\bibfnamefont {W.~F.}\ \bibnamefont {Koehl}}, \bibinfo {author}
  {\bibfnamefont {T.}~\bibnamefont {Ohshima}}, \bibinfo {author} {\bibfnamefont
  {N.~T.}\ \bibnamefont {Son}}, \bibinfo {author} {\bibfnamefont
  {E.}~\bibnamefont {Janz{\'e}n}}, \bibinfo {author} {\bibfnamefont {A.}\
  \bibnamefont {Gali}},\ and\ \bibinfo {author} {\bibfnamefont {D.~D.}\
  \bibnamefont {Awschalom}},\ }\bibfield  {title} {\bibinfo {title} {{Isolated
  Spin Qubits in SiC with a High-Fidelity Infrared Spin-to-Photon Interface}},\
  }\href@noop {} {\bibfield  {journal} {\bibinfo  {journal} {Physical Review
  X}\ }\textbf {\bibinfo {volume} {7}},\ \bibinfo {pages} {021046} (\bibinfo
  {year} {2017})}\BibitemShut {NoStop}%
\bibitem [{\citenamefont {Nagy}\ \emph {et~al.}(2019)\citenamefont {Nagy},
  \citenamefont {Niethammer}, \citenamefont {Widmann}, \citenamefont {Chen},
  \citenamefont {Udvarhelyi}, \citenamefont {Bonato}, \citenamefont {Hassan},
  \citenamefont {Karhu}, \citenamefont {Ivanov}, \citenamefont {Son},
  \citenamefont {Maze}, \citenamefont {Ohshima}, \citenamefont {Soykal},
  \citenamefont {Gali}, \citenamefont {Lee}, \citenamefont {Kaiser},\ and\
  \citenamefont {Wrachtrup}}]{Nagy:2019fw}%
  \BibitemOpen
  \bibfield  {author} {\bibinfo {author} {\bibfnamefont {R.}~\bibnamefont
  {Nagy}}, \bibinfo {author} {\bibfnamefont {M.}~\bibnamefont {Niethammer}},
  \bibinfo {author} {\bibfnamefont {M.}~\bibnamefont {Widmann}}, \bibinfo
  {author} {\bibfnamefont {Y.-C.}\ \bibnamefont {Chen}}, \bibinfo {author}
  {\bibfnamefont {P.}~\bibnamefont {Udvarhelyi}}, \bibinfo {author}
  {\bibfnamefont {C.}~\bibnamefont {Bonato}}, \bibinfo {author} {\bibfnamefont
  {J.~u.}\ \bibnamefont {Hassan}}, \bibinfo {author} {\bibfnamefont
  {R.}~\bibnamefont {Karhu}}, \bibinfo {author} {\bibfnamefont {I.~G.}\
  \bibnamefont {Ivanov}}, \bibinfo {author} {\bibfnamefont {N.~T.}\
  \bibnamefont {Son}}, \bibinfo {author} {\bibfnamefont {J.~R.}\ \bibnamefont
  {Maze}}, \bibinfo {author} {\bibfnamefont {T.}~\bibnamefont {Ohshima}},
  \bibinfo {author} {\bibfnamefont {{\"O}.~O.}\ \bibnamefont {Soykal}},
  \bibinfo {author} {\bibfnamefont {A.}~\bibnamefont {Gali}}, \bibinfo {author}
  {\bibfnamefont {S.-Y.}\ \bibnamefont {Lee}}, \bibinfo {author} {\bibfnamefont
  {F.}~\bibnamefont {Kaiser}},\ and\ \bibinfo {author} {\bibfnamefont
  {J.}~\bibnamefont {Wrachtrup}},\ }\bibfield  {title} {\bibinfo {title}
  {{High-fidelity spin and optical control of single silicon-vacancy centres in
  silicon carbide}},\ }\href@noop {} {\bibfield  {journal} {\bibinfo  {journal}
  {Nature Communications}\ }\textbf {\bibinfo {volume} {10}},\ \bibinfo {pages}
  {1954} (\bibinfo {year} {2019})}\BibitemShut {NoStop}%
\bibitem [{\citenamefont {Udvarhelyi}\ \emph {et~al.}(2019)\citenamefont
  {Udvarhelyi}, \citenamefont {Nagy}, \citenamefont {Kaiser}, \citenamefont
  {Lee}, \citenamefont {Wrachtrup},\ and\ \citenamefont
  {Gali}}]{Udvarhelyi:2019eh}%
  \BibitemOpen
  \bibfield  {author} {\bibinfo {author} {\bibfnamefont {P.}~\bibnamefont
  {Udvarhelyi}}, \bibinfo {author} {\bibfnamefont {R.}~\bibnamefont {Nagy}},
  \bibinfo {author} {\bibfnamefont {F.}~\bibnamefont {Kaiser}}, \bibinfo
  {author} {\bibfnamefont {S.-Y.}\ \bibnamefont {Lee}}, \bibinfo {author}
  {\bibfnamefont {J.}~\bibnamefont {Wrachtrup}},\ and\ \bibinfo {author}
  {\bibfnamefont {A.}~\bibnamefont {Gali}},\ }\bibfield  {title} {\bibinfo
  {title} {{Spectrally Stable Defect Qubits with no Inversion Symmetry for
  Robust Spin-To-Photon Interface}},\ }\href@noop {} {\bibfield  {journal}
  {\bibinfo  {journal} {Physical Review Applied}\ }\textbf {\bibinfo {volume}
  {11}},\ \bibinfo {pages} {044022} (\bibinfo {year} {2019})}\BibitemShut
  {NoStop}%
\bibitem [{\citenamefont {Kraus}\ \emph {et~al.}(2017)\citenamefont {Kraus},
  \citenamefont {Simin}, \citenamefont {Kasper}, \citenamefont {Suda},
  \citenamefont {Kawabata}, \citenamefont {Kada}, \citenamefont {Honda},
  \citenamefont {Hijikata}, \citenamefont {Ohshima}, \citenamefont {Dyakonov},\
  and\ \citenamefont {Astakhov}}]{Kraus:2017cka}%
  \BibitemOpen
  \bibfield  {author} {\bibinfo {author} {\bibfnamefont {H.}~\bibnamefont
  {Kraus}}, \bibinfo {author} {\bibfnamefont {D.}~\bibnamefont {Simin}},
  \bibinfo {author} {\bibfnamefont {C.}~\bibnamefont {Kasper}}, \bibinfo
  {author} {\bibfnamefont {Y.}~\bibnamefont {Suda}}, \bibinfo {author}
  {\bibfnamefont {S.}~\bibnamefont {Kawabata}}, \bibinfo {author}
  {\bibfnamefont {W.}~\bibnamefont {Kada}}, \bibinfo {author} {\bibfnamefont
  {T.}~\bibnamefont {Honda}}, \bibinfo {author} {\bibfnamefont
  {Y.}~\bibnamefont {Hijikata}}, \bibinfo {author} {\bibfnamefont
  {T.}~\bibnamefont {Ohshima}}, \bibinfo {author} {\bibfnamefont
  {V.}~\bibnamefont {Dyakonov}},\ and\ \bibinfo {author} {\bibfnamefont
  {G.~V.}\ \bibnamefont {Astakhov}},\ }\bibfield  {title} {\bibinfo {title}
  {{Three-Dimensional Proton Beam Writing of Optically Active Coherent Vacancy
  Spins in Silicon Carbide}},\ }\href@noop {} {\bibfield  {journal} {\bibinfo
  {journal} {Nano Letters}\ }\textbf {\bibinfo {volume} {17}},\ \bibinfo
  {pages} {2865} (\bibinfo {year} {2017})}\BibitemShut {NoStop}%
\bibitem [{\citenamefont {Wang}\ \emph {et~al.}(2017)\citenamefont {Wang},
  \citenamefont {Zhang}, \citenamefont {Zhou}, \citenamefont {Li},
  \citenamefont {Wang}, \citenamefont {Peddibhotla}, \citenamefont {Liu},
  \citenamefont {Bauerdick}, \citenamefont {Rudzinski}, \citenamefont {Liu},\
  and\ \citenamefont {Gao}}]{Wang:2017fb}%
  \BibitemOpen
  \bibfield  {author} {\bibinfo {author} {\bibfnamefont {J.}~\bibnamefont
  {Wang}}, \bibinfo {author} {\bibfnamefont {X.}~\bibnamefont {Zhang}},
  \bibinfo {author} {\bibfnamefont {Y.}~\bibnamefont {Zhou}}, \bibinfo {author}
  {\bibfnamefont {K.}~\bibnamefont {Li}}, \bibinfo {author} {\bibfnamefont
  {Z.}~\bibnamefont {Wang}}, \bibinfo {author} {\bibfnamefont {P.}~\bibnamefont
  {Peddibhotla}}, \bibinfo {author} {\bibfnamefont {F.}~\bibnamefont {Liu}},
  \bibinfo {author} {\bibfnamefont {S.}~\bibnamefont {Bauerdick}}, \bibinfo
  {author} {\bibfnamefont {A.}~\bibnamefont {Rudzinski}}, \bibinfo {author}
  {\bibfnamefont {Z.}~\bibnamefont {Liu}},\ and\ \bibinfo {author}
  {\bibfnamefont {W.}~\bibnamefont {Gao}},\ }\bibfield  {title} {\bibinfo
  {title} {{Scalable Fabrication of Single Silicon Vacancy Defect Arrays in
  Silicon Carbide Using Focused Ion Beam}},\ }\href@noop {} {\bibfield
  {journal} {\bibinfo  {journal} {ACS Photonics}\ }\textbf {\bibinfo {volume}
  {4}},\ \bibinfo {pages} {1054} (\bibinfo {year} {2017})}\BibitemShut
  {NoStop}%
\bibitem [{\citenamefont {Wang}\ \emph {et~al.}(2019)\citenamefont {Wang},
  \citenamefont {Li}, \citenamefont {Yan}, \citenamefont {Liu}, \citenamefont
  {Guo}, \citenamefont {Zhang}, \citenamefont {Zhou}, \citenamefont {Guo},
  \citenamefont {Lin}, \citenamefont {Cui}, \citenamefont {Xu}, \citenamefont
  {Xu}, \citenamefont {Li},\ and\ \citenamefont {Guo}}]{Wang:2019dk}%
  \BibitemOpen
  \bibfield  {author} {\bibinfo {author} {\bibfnamefont {J.-F.}\ \bibnamefont
  {Wang}}, \bibinfo {author} {\bibfnamefont {Q.}~\bibnamefont {Li}}, \bibinfo
  {author} {\bibfnamefont {F.-F.}\ \bibnamefont {Yan}}, \bibinfo {author}
  {\bibfnamefont {H.}~\bibnamefont {Liu}}, \bibinfo {author} {\bibfnamefont
  {G.-P.}\ \bibnamefont {Guo}}, \bibinfo {author} {\bibfnamefont {W.-P.}\
  \bibnamefont {Zhang}}, \bibinfo {author} {\bibfnamefont {X.}~\bibnamefont
  {Zhou}}, \bibinfo {author} {\bibfnamefont {L.-P.}\ \bibnamefont {Guo}},
  \bibinfo {author} {\bibfnamefont {Z.-H.}\ \bibnamefont {Lin}}, \bibinfo
  {author} {\bibfnamefont {J.-M.}\ \bibnamefont {Cui}}, \bibinfo {author}
  {\bibfnamefont {X.-Y.}\ \bibnamefont {Xu}}, \bibinfo {author} {\bibfnamefont
  {J.-S.}\ \bibnamefont {Xu}}, \bibinfo {author} {\bibfnamefont {C.-F.}\
  \bibnamefont {Li}},\ and\ \bibinfo {author} {\bibfnamefont {G.-C.}\
  \bibnamefont {Guo}},\ }\bibfield  {title} {\bibinfo {title} {{On-Demand
  Generation of Single Silicon Vacancy Defects in Silicon Carbide}},\
  }\href@noop {} {\bibfield  {journal} {\bibinfo  {journal} {ACS Photonics}\
  }\textbf {\bibinfo {volume} {6}},\ \bibinfo {pages} {1736} (\bibinfo {year}
  {2019})}\BibitemShut {NoStop}%
\bibitem [{\citenamefont {Fuchs}\ \emph {et~al.}(2013)\citenamefont {Fuchs},
  \citenamefont {Soltamov}, \citenamefont {V{\"a}th}, \citenamefont {Baranov},
  \citenamefont {Mokhov}, \citenamefont {Astakhov},\ and\ \citenamefont
  {Dyakonov}}]{Fuchs:2013dz}%
  \BibitemOpen
  \bibfield  {author} {\bibinfo {author} {\bibfnamefont {F.}~\bibnamefont
  {Fuchs}}, \bibinfo {author} {\bibfnamefont {V.~A.}\ \bibnamefont {Soltamov}},
  \bibinfo {author} {\bibfnamefont {S.}~\bibnamefont {V{\"a}th}}, \bibinfo
  {author} {\bibfnamefont {P.~G.}\ \bibnamefont {Baranov}}, \bibinfo {author}
  {\bibfnamefont {E.~N.}\ \bibnamefont {Mokhov}}, \bibinfo {author}
  {\bibfnamefont {G.~V.}\ \bibnamefont {Astakhov}},\ and\ \bibinfo {author}
  {\bibfnamefont {V.}~\bibnamefont {Dyakonov}},\ }\bibfield  {title} {\bibinfo
  {title} {{Silicon carbide light-emitting diode as a prospective room
  temperature source for single photons}},\ }\href@noop {} {\bibfield
  {journal} {\bibinfo  {journal} {Scientific Reports}\ }\textbf {\bibinfo
  {volume} {3}},\ \bibinfo {pages} {1637} (\bibinfo {year} {2013})}\BibitemShut
  {NoStop}%
\bibitem [{\citenamefont {Lohrmann}\ \emph {et~al.}(2015)\citenamefont
  {Lohrmann}, \citenamefont {Iwamoto}, \citenamefont {Bodrog}, \citenamefont
  {Castelletto}, \citenamefont {Ohshima}, \citenamefont {Karle}, \citenamefont
  {Gali}, \citenamefont {Prawer}, \citenamefont {McCallum},\ and\ \citenamefont
  {Johnson}}]{Lohrmann:2015hd}%
  \BibitemOpen
  \bibfield  {author} {\bibinfo {author} {\bibfnamefont {A.}~\bibnamefont
  {Lohrmann}}, \bibinfo {author} {\bibfnamefont {N.}~\bibnamefont {Iwamoto}},
  \bibinfo {author} {\bibfnamefont {Z.}~\bibnamefont {Bodrog}}, \bibinfo
  {author} {\bibfnamefont {S.}~\bibnamefont {Castelletto}}, \bibinfo {author}
  {\bibfnamefont {T.}~\bibnamefont {Ohshima}}, \bibinfo {author} {\bibfnamefont
  {T.~J.}\ \bibnamefont {Karle}}, \bibinfo {author} {\bibfnamefont
  {A.}~\bibnamefont {Gali}}, \bibinfo {author} {\bibfnamefont {S.}~\bibnamefont
  {Prawer}}, \bibinfo {author} {\bibfnamefont {J.~C.}\ \bibnamefont
  {McCallum}},\ and\ \bibinfo {author} {\bibfnamefont {B.~C.}\ \bibnamefont
  {Johnson}},\ }\bibfield  {title} {\bibinfo {title} {{Single-photon emitting
  diode in silicon carbide}},\ }\href@noop {} {\bibfield  {journal} {\bibinfo
  {journal} {Nature Communications}\ }\textbf {\bibinfo {volume} {6}},\
  \bibinfo {pages} {7783} (\bibinfo {year} {2015})}\BibitemShut {NoStop}%
\bibitem [{\citenamefont {Sato}\ \emph {et~al.}(2018)\citenamefont {Sato},
  \citenamefont {Honda}, \citenamefont {Makino}, \citenamefont {Hijikata},
  \citenamefont {Lee},\ and\ \citenamefont {Ohshima}}]{Sato:2018jq}%
  \BibitemOpen
  \bibfield  {author} {\bibinfo {author} {\bibfnamefont {S.-i.}\ \bibnamefont
  {Sato}}, \bibinfo {author} {\bibfnamefont {T.}~\bibnamefont {Honda}},
  \bibinfo {author} {\bibfnamefont {T.}~\bibnamefont {Makino}}, \bibinfo
  {author} {\bibfnamefont {Y.}~\bibnamefont {Hijikata}}, \bibinfo {author}
  {\bibfnamefont {S.-Y.}\ \bibnamefont {Lee}},\ and\ \bibinfo {author}
  {\bibfnamefont {T.}~\bibnamefont {Ohshima}},\ }\bibfield  {title} {\bibinfo
  {title} {{Room Temperature Electrical Control of Single Photon Sources at
  4H-SiC Surface}},\ }\href@noop {} {\bibfield  {journal} {\bibinfo  {journal}
  {ACS Photonics}\ ,\ \bibinfo {pages} {acsphotonics.8b00375}} (\bibinfo {year}
  {2018})}\BibitemShut {NoStop}%
\bibitem [{\citenamefont {Widmann}\ \emph {et~al.}(2018)\citenamefont
  {Widmann}, \citenamefont {Niethammer}, \citenamefont {Makino}, \citenamefont
  {Rendler}, \citenamefont {Lasse}, \citenamefont {Ohshima}, \citenamefont
  {Ul~Hassan}, \citenamefont {Tien~Son}, \citenamefont {Lee},\ and\
  \citenamefont {Wrachtrup}}]{Widmann:2018jh}%
  \BibitemOpen
  \bibfield  {author} {\bibinfo {author} {\bibfnamefont {M.}~\bibnamefont
  {Widmann}}, \bibinfo {author} {\bibfnamefont {M.}~\bibnamefont {Niethammer}},
  \bibinfo {author} {\bibfnamefont {T.}~\bibnamefont {Makino}}, \bibinfo
  {author} {\bibfnamefont {T.}~\bibnamefont {Rendler}}, \bibinfo {author}
  {\bibfnamefont {S.}~\bibnamefont {Lasse}}, \bibinfo {author} {\bibfnamefont
  {T.}~\bibnamefont {Ohshima}}, \bibinfo {author} {\bibfnamefont
  {J.}~\bibnamefont {Ul~Hassan}}, \bibinfo {author} {\bibfnamefont
  {N.}~\bibnamefont {Tien~Son}}, \bibinfo {author} {\bibfnamefont {S.-Y.}\
  \bibnamefont {Lee}},\ and\ \bibinfo {author} {\bibfnamefont {J.}~\bibnamefont
  {Wrachtrup}},\ }\bibfield  {title} {\bibinfo {title} {{Bright single photon
  sources in lateral silicon carbide light emitting diodes}},\ }\href@noop {}
  {\bibfield  {journal} {\bibinfo  {journal} {Applied Physics Letters}\
  }\textbf {\bibinfo {volume} {112}},\ \bibinfo {pages} {231103} (\bibinfo
  {year} {2018})}\BibitemShut {NoStop}%
\bibitem [{\citenamefont {Klimov}\ \emph {et~al.}(2014)\citenamefont {Klimov},
  \citenamefont {Falk}, \citenamefont {Buckley},\ and\ \citenamefont
  {Awschalom}}]{Klimov:2013ua}%
  \BibitemOpen
  \bibfield  {author} {\bibinfo {author} {\bibfnamefont {P.~V.}\ \bibnamefont
  {Klimov}}, \bibinfo {author} {\bibfnamefont {A.~L.}\ \bibnamefont {Falk}},
  \bibinfo {author} {\bibfnamefont {B.~B.}\ \bibnamefont {Buckley}},\ and\
  \bibinfo {author} {\bibfnamefont {D.~D.}\ \bibnamefont {Awschalom}},\
  }\bibfield  {title} {\bibinfo {title} {{Electrically Driven Spin Resonance in
  Silicon Carbide Color Centers}},\ }\href@noop {} {\bibfield  {journal}
  {\bibinfo  {journal} {Physical Review Letters}\ }\textbf {\bibinfo {volume}
  {112}},\ \bibinfo {pages} {087601} (\bibinfo {year} {2014})}\BibitemShut
  {NoStop}%
\bibitem [{\citenamefont {Falk}\ \emph {et~al.}(2014)\citenamefont {Falk},
  \citenamefont {Klimov}, \citenamefont {Buckley}, \citenamefont {Iv{\'a}dy},
  \citenamefont {Abrikosov}, \citenamefont {Calusine}, \citenamefont {Koehl},
  \citenamefont {Gali},\ and\ \citenamefont {Awschalom}}]{Falk:2014fh}%
  \BibitemOpen
  \bibfield  {author} {\bibinfo {author} {\bibfnamefont {A.~L.}\ \bibnamefont
  {Falk}}, \bibinfo {author} {\bibfnamefont {P.~V.}\ \bibnamefont {Klimov}},
  \bibinfo {author} {\bibfnamefont {B.~B.}\ \bibnamefont {Buckley}}, \bibinfo
  {author} {\bibfnamefont {V.}~\bibnamefont {Iv{\'a}dy}}, \bibinfo {author}
  {\bibfnamefont {I.~A.}\ \bibnamefont {Abrikosov}}, \bibinfo {author}
  {\bibfnamefont {G.}~\bibnamefont {Calusine}}, \bibinfo {author}
  {\bibfnamefont {W.~F.}\ \bibnamefont {Koehl}}, \bibinfo {author}
  {\bibfnamefont {A.}~\bibnamefont {Gali}},\ and\ \bibinfo {author}
  {\bibfnamefont {D.~D.}\ \bibnamefont {Awschalom}},\ }\bibfield  {title}
  {\bibinfo {title} {{Electrically and Mechanically Tunable Electron Spins in
  Silicon Carbide Color Centers}},\ }\href@noop {} {\bibfield  {journal}
  {\bibinfo  {journal} {Physical Review Letters}\ }\textbf {\bibinfo {volume}
  {112}},\ \bibinfo {pages} {187601} (\bibinfo {year} {2014})}\BibitemShut
  {NoStop}%
\bibitem [{\citenamefont {Widmann}\ \emph {et~al.}(2019)\citenamefont
  {Widmann}, \citenamefont {Niethammer}, \citenamefont {Fedyanin},
  \citenamefont {Khramtsov}, \citenamefont {Rendler}, \citenamefont {Booker},
  \citenamefont {Ul~Hassan}, \citenamefont {Morioka}, \citenamefont {Chen},
  \citenamefont {Ivanov}, \citenamefont {Son}, \citenamefont {Ohshima},
  \citenamefont {Bockstedte}, \citenamefont {Gali}, \citenamefont {Bonato},
  \citenamefont {Lee},\ and\ \citenamefont {Wrachtrup}}]{Widmann:2019ja}%
  \BibitemOpen
  \bibfield  {author} {\bibinfo {author} {\bibfnamefont {M.}~\bibnamefont
  {Widmann}}, \bibinfo {author} {\bibfnamefont {M.}~\bibnamefont {Niethammer}},
  \bibinfo {author} {\bibfnamefont {D.~Y.}\ \bibnamefont {Fedyanin}}, \bibinfo
  {author} {\bibfnamefont {I.~A.}\ \bibnamefont {Khramtsov}}, \bibinfo {author}
  {\bibfnamefont {T.}~\bibnamefont {Rendler}}, \bibinfo {author} {\bibfnamefont
  {I.~D.}\ \bibnamefont {Booker}}, \bibinfo {author} {\bibfnamefont
  {J.}~\bibnamefont {Ul~Hassan}}, \bibinfo {author} {\bibfnamefont
  {N.}~\bibnamefont {Morioka}}, \bibinfo {author} {\bibfnamefont {Y.-C.}\
  \bibnamefont {Chen}}, \bibinfo {author} {\bibfnamefont {I.~G.}\ \bibnamefont
  {Ivanov}}, \bibinfo {author} {\bibfnamefont {N.~T.}\ \bibnamefont {Son}},
  \bibinfo {author} {\bibfnamefont {T.}~\bibnamefont {Ohshima}}, \bibinfo
  {author} {\bibfnamefont {M.}~\bibnamefont {Bockstedte}}, \bibinfo {author}
  {\bibfnamefont {A.}~\bibnamefont {Gali}}, \bibinfo {author} {\bibfnamefont
  {C.}~\bibnamefont {Bonato}}, \bibinfo {author} {\bibfnamefont {S.-Y.}\
  \bibnamefont {Lee}},\ and\ \bibinfo {author} {\bibfnamefont {J.}~\bibnamefont
  {Wrachtrup}},\ }\bibfield  {title} {\bibinfo {title} {{Electrical Charge
  State Manipulation of Single Silicon Vacancies in a Silicon Carbide Quantum
  Optoelectronic Device}},\ }\href@noop {} {\bibfield  {journal} {\bibinfo
  {journal} {Nano Letters}\ }\textbf {\bibinfo {volume} {19}},\ \bibinfo
  {pages} {7173} (\bibinfo {year} {2019})}\BibitemShut {NoStop}%
\bibitem [{\citenamefont {Whiteley}\ \emph {et~al.}(2019)\citenamefont
  {Whiteley}, \citenamefont {Wolfowicz}, \citenamefont {Anderson},
  \citenamefont {Bourassa}, \citenamefont {Ma}, \citenamefont {Ye},
  \citenamefont {Koolstra}, \citenamefont {Satzinger}, \citenamefont {Holt},
  \citenamefont {Heremans}, \citenamefont {Cleland}, \citenamefont {Schuster},
  \citenamefont {Galli},\ and\ \citenamefont {Awschalom}}]{Whiteley:2019eu}%
  \BibitemOpen
  \bibfield  {author} {\bibinfo {author} {\bibfnamefont {S.~J.}\ \bibnamefont
  {Whiteley}}, \bibinfo {author} {\bibfnamefont {G.}~\bibnamefont {Wolfowicz}},
  \bibinfo {author} {\bibfnamefont {C.~P.}\ \bibnamefont {Anderson}}, \bibinfo
  {author} {\bibfnamefont {A.}~\bibnamefont {Bourassa}}, \bibinfo {author}
  {\bibfnamefont {H.}~\bibnamefont {Ma}}, \bibinfo {author} {\bibfnamefont
  {M.}~\bibnamefont {Ye}}, \bibinfo {author} {\bibfnamefont {G.}~\bibnamefont
  {Koolstra}}, \bibinfo {author} {\bibfnamefont {K.~J.}\ \bibnamefont
  {Satzinger}}, \bibinfo {author} {\bibfnamefont {M.~V.}\ \bibnamefont {Holt}},
  \bibinfo {author} {\bibfnamefont {F.~J.}\ \bibnamefont {Heremans}}, \bibinfo
  {author} {\bibfnamefont {A.~N.}\ \bibnamefont {Cleland}}, \bibinfo {author}
  {\bibfnamefont {D.~I.}\ \bibnamefont {Schuster}}, \bibinfo {author}
  {\bibfnamefont {G.}~\bibnamefont {Galli}},\ and\ \bibinfo {author}
  {\bibfnamefont {D.~D.}\ \bibnamefont {Awschalom}},\ }\bibfield  {title}
  {\bibinfo {title} {{Spin{\textendash}phonon interactions in silicon carbide
  addressed by Gaussian acoustics}},\ }\href@noop {} {\bibfield  {journal}
  {\bibinfo  {journal} {Nature Physics}\ }\textbf {\bibinfo {volume} {15}},\
  \bibinfo {pages} {490} (\bibinfo {year} {2019})}\BibitemShut {NoStop}%
\bibitem [{\citenamefont {Poshakinskiy}\ and\ \citenamefont
  {Astakhov}(2019)}]{Poshakinskiy:2019bi}%
  \BibitemOpen
  \bibfield  {author} {\bibinfo {author} {\bibfnamefont {A.~V.}\ \bibnamefont
  {Poshakinskiy}}\ and\ \bibinfo {author} {\bibfnamefont {G.~V.}\ \bibnamefont
  {Astakhov}},\ }\bibfield  {title} {\bibinfo {title} {{Optically detected
  spin-mechanical resonance in silicon carbide membranes}},\ }\href@noop {}
  {\bibfield  {journal} {\bibinfo  {journal} {Physical Review B}\ }\textbf
  {\bibinfo {volume} {100}},\ \bibinfo {pages} {094104} (\bibinfo {year}
  {2019})}\BibitemShut {NoStop}%
\bibitem [{\citenamefont {Castelletto}\ \emph {et~al.}(2014)\citenamefont
  {Castelletto}, \citenamefont {Johnson}, \citenamefont {Zachreson},
  \citenamefont {Beke}, \citenamefont {Balogh}, \citenamefont {Ohshima},
  \citenamefont {Aharonovich},\ and\ \citenamefont
  {Gali}}]{Castelletto:2014eu}%
  \BibitemOpen
  \bibfield  {author} {\bibinfo {author} {\bibfnamefont {S.}~\bibnamefont
  {Castelletto}}, \bibinfo {author} {\bibfnamefont {B.~C.}\ \bibnamefont
  {Johnson}}, \bibinfo {author} {\bibfnamefont {C.}~\bibnamefont {Zachreson}},
  \bibinfo {author} {\bibfnamefont {D.}~\bibnamefont {Beke}}, \bibinfo {author}
  {\bibfnamefont {I.}~\bibnamefont {Balogh}}, \bibinfo {author} {\bibfnamefont
  {T.}~\bibnamefont {Ohshima}}, \bibinfo {author} {\bibfnamefont
  {I.}~\bibnamefont {Aharonovich}},\ and\ \bibinfo {author} {\bibfnamefont
  {A.}~\bibnamefont {Gali}},\ }\bibfield  {title} {\bibinfo {title} {{Room
  Temperature Quantum Emission from Cubic Silicon Carbide Nanoparticles}},\
  }\href@noop {} {\bibfield  {journal} {\bibinfo  {journal} {ACS Nano}\
  }\textbf {\bibinfo {volume} {8}},\ \bibinfo {pages} {7938} (\bibinfo {year}
  {2014})}\BibitemShut {NoStop}%
\bibitem [{\citenamefont {Muzha}\ \emph {et~al.}(2014)\citenamefont {Muzha},
  \citenamefont {Fuchs}, \citenamefont {Tarakina}, \citenamefont {Simin},
  \citenamefont {Trupke}, \citenamefont {Soltamov}, \citenamefont {Mokhov},
  \citenamefont {Baranov}, \citenamefont {Dyakonov}, \citenamefont {Krueger},\
  and\ \citenamefont {Astakhov}}]{Muzha:2014th}%
  \BibitemOpen
  \bibfield  {author} {\bibinfo {author} {\bibfnamefont {A.}~\bibnamefont
  {Muzha}}, \bibinfo {author} {\bibfnamefont {F.}~\bibnamefont {Fuchs}},
  \bibinfo {author} {\bibfnamefont {N.~V.}\ \bibnamefont {Tarakina}}, \bibinfo
  {author} {\bibfnamefont {D.}~\bibnamefont {Simin}}, \bibinfo {author}
  {\bibfnamefont {M.}~\bibnamefont {Trupke}}, \bibinfo {author} {\bibfnamefont
  {V.~A.}\ \bibnamefont {Soltamov}}, \bibinfo {author} {\bibfnamefont {E.~N.}\
  \bibnamefont {Mokhov}}, \bibinfo {author} {\bibfnamefont {P.~G.}\
  \bibnamefont {Baranov}}, \bibinfo {author} {\bibfnamefont {V.}~\bibnamefont
  {Dyakonov}}, \bibinfo {author} {\bibfnamefont {A.}~\bibnamefont {Krueger}},\
  and\ \bibinfo {author} {\bibfnamefont {G.~V.}\ \bibnamefont {Astakhov}},\
  }\bibfield  {title} {\bibinfo {title} {{Room-temperature near-infrared
  silicon carbide nanocrystalline emitters based on optically aligned spin
  defects}},\ }\href@noop {} {\bibfield  {journal} {\bibinfo  {journal}
  {Applied Physics Letters}\ }\textbf {\bibinfo {volume} {105}},\ \bibinfo
  {pages} {243112} (\bibinfo {year} {2014})}\BibitemShut {NoStop}%
\bibitem [{\citenamefont {S{\"o}rman}\ \emph {et~al.}(2000)\citenamefont
  {S{\"o}rman}, \citenamefont {Son}, \citenamefont {Chen}, \citenamefont
  {Kordina}, \citenamefont {Hallin},\ and\ \citenamefont
  {Janz{\'e}n}}]{Sorman:2000ij}%
  \BibitemOpen
  \bibfield  {author} {\bibinfo {author} {\bibfnamefont {E.}~\bibnamefont
  {S{\"o}rman}}, \bibinfo {author} {\bibfnamefont {N.}~\bibnamefont {Son}},
  \bibinfo {author} {\bibfnamefont {W.}~\bibnamefont {Chen}}, \bibinfo {author}
  {\bibfnamefont {O.}~\bibnamefont {Kordina}}, \bibinfo {author} {\bibfnamefont
  {C.}~\bibnamefont {Hallin}},\ and\ \bibinfo {author} {\bibfnamefont
  {E.}~\bibnamefont {Janz{\'e}n}},\ }\bibfield  {title} {\bibinfo {title}
  {{Silicon vacancy related defect in 4H and 6H SiC}},\ }\href@noop {}
  {\bibfield  {journal} {\bibinfo  {journal} {Physical Review B}\ }\textbf
  {\bibinfo {volume} {61}},\ \bibinfo {pages} {2613} (\bibinfo {year}
  {2000})}\BibitemShut {NoStop}%
\bibitem [{\citenamefont {Son}\ \emph {et~al.}(2006)\citenamefont {Son},
  \citenamefont {Carlsson}, \citenamefont {ul~Hassan}, \citenamefont
  {Janz{\'e}n}, \citenamefont {Umeda}, \citenamefont {Isoya}, \citenamefont
  {Gali}, \citenamefont {Bockstedte}, \citenamefont {Morishita}, \citenamefont
  {Ohshima},\ and\ \citenamefont {Itoh}}]{Son:2006im}%
  \BibitemOpen
  \bibfield  {author} {\bibinfo {author} {\bibfnamefont {N.}~\bibnamefont
  {Son}}, \bibinfo {author} {\bibfnamefont {P.}~\bibnamefont {Carlsson}},
  \bibinfo {author} {\bibfnamefont {J.}~\bibnamefont {ul~Hassan}}, \bibinfo
  {author} {\bibfnamefont {E.}~\bibnamefont {Janz{\'e}n}}, \bibinfo {author}
  {\bibfnamefont {T.}~\bibnamefont {Umeda}}, \bibinfo {author} {\bibfnamefont
  {J.}~\bibnamefont {Isoya}}, \bibinfo {author} {\bibfnamefont
  {A.}~\bibnamefont {Gali}}, \bibinfo {author} {\bibfnamefont {M.}~\bibnamefont
  {Bockstedte}}, \bibinfo {author} {\bibfnamefont {N.}~\bibnamefont
  {Morishita}}, \bibinfo {author} {\bibfnamefont {T.}~\bibnamefont {Ohshima}},\
  and\ \bibinfo {author} {\bibfnamefont {H.}~\bibnamefont {Itoh}},\ }\bibfield
  {title} {\bibinfo {title} {{Divacancy in 4H-SiC}},\ }\href@noop {} {\bibfield
   {journal} {\bibinfo  {journal} {Physical Review Letters}\ }\textbf {\bibinfo
  {volume} {96}},\ \bibinfo {pages} {055501} (\bibinfo {year}
  {2006})}\BibitemShut {NoStop}%
\bibitem [{\citenamefont {Gali}\ \emph {et~al.}(2011)\citenamefont {Gali},
  \citenamefont {Simon},\ and\ \citenamefont {Lowther}}]{Gali:2011fn}%
  \BibitemOpen
  \bibfield  {author} {\bibinfo {author} {\bibfnamefont {A.}~\bibnamefont
  {Gali}}, \bibinfo {author} {\bibfnamefont {T.}~\bibnamefont {Simon}},\ and\
  \bibinfo {author} {\bibfnamefont {J.~E.}\ \bibnamefont {Lowther}},\
  }\bibfield  {title} {\bibinfo {title} {{An ab initio study of local vibration
  modes of the nitrogen-vacancy center in diamond}},\ }\href@noop {} {\bibfield
   {journal} {\bibinfo  {journal} {New Journal of Physics}\ }\textbf {\bibinfo
  {volume} {13}},\ \bibinfo {pages} {025016} (\bibinfo {year}
  {2011})}\BibitemShut {NoStop}%
\bibitem [{\citenamefont {Alkauskas}\ \emph {et~al.}(2014)\citenamefont
  {Alkauskas}, \citenamefont {Buckley}, \citenamefont {Awschalom},\ and\
  \citenamefont {Van~de Walle}}]{Alkauskas:2014gq}%
  \BibitemOpen
  \bibfield  {author} {\bibinfo {author} {\bibfnamefont {A.}~\bibnamefont
  {Alkauskas}}, \bibinfo {author} {\bibfnamefont {B.~B.}\ \bibnamefont
  {Buckley}}, \bibinfo {author} {\bibfnamefont {D.~D.}\ \bibnamefont
  {Awschalom}},\ and\ \bibinfo {author} {\bibfnamefont {C.~G.}\ \bibnamefont
  {Van~de Walle}},\ }\bibfield  {title} {\bibinfo {title} {{First-principles
  theory of the luminescence lineshape for the triplet transition in diamond NV
  centres}},\ }\href@noop {} {\bibfield  {journal} {\bibinfo  {journal} {New
  Journal of Physics}\ }\textbf {\bibinfo {volume} {16}},\ \bibinfo {pages}
  {073026} (\bibinfo {year} {2014})}\BibitemShut {NoStop}%
\bibitem [{\citenamefont {Nagy}\ \emph {et~al.}(2018)\citenamefont {Nagy},
  \citenamefont {Widmann}, \citenamefont {Niethammer}, \citenamefont {Dasari},
  \citenamefont {Gerhardt}, \citenamefont {Soykal}, \citenamefont {Radulaski},
  \citenamefont {Ohshima}, \citenamefont {Vu{\v c}kovi{\'c}}, \citenamefont
  {Son}, \citenamefont {Ivanov}, \citenamefont {Economou}, \citenamefont
  {Bonato}, \citenamefont {Lee},\ and\ \citenamefont
  {Wrachtrup}}]{Nagy:2018ey}%
  \BibitemOpen
  \bibfield  {author} {\bibinfo {author} {\bibfnamefont {R.}~\bibnamefont
  {Nagy}}, \bibinfo {author} {\bibfnamefont {M.}~\bibnamefont {Widmann}},
  \bibinfo {author} {\bibfnamefont {M.}~\bibnamefont {Niethammer}}, \bibinfo
  {author} {\bibfnamefont {D.~B.~R.}\ \bibnamefont {Dasari}}, \bibinfo {author}
  {\bibfnamefont {I.}~\bibnamefont {Gerhardt}}, \bibinfo {author}
  {\bibfnamefont {{\"O}.~O.}\ \bibnamefont {Soykal}}, \bibinfo {author}
  {\bibfnamefont {M.}~\bibnamefont {Radulaski}}, \bibinfo {author}
  {\bibfnamefont {T.}~\bibnamefont {Ohshima}}, \bibinfo {author} {\bibfnamefont
  {J.}~\bibnamefont {Vu{\v c}kovi{\'c}}}, \bibinfo {author} {\bibfnamefont
  {N.~T.}\ \bibnamefont {Son}}, \bibinfo {author} {\bibfnamefont {I.~G.}\
  \bibnamefont {Ivanov}}, \bibinfo {author} {\bibfnamefont {S.~E.}\
  \bibnamefont {Economou}}, \bibinfo {author} {\bibfnamefont {C.}~\bibnamefont
  {Bonato}}, \bibinfo {author} {\bibfnamefont {S.-Y.}\ \bibnamefont {Lee}},\
  and\ \bibinfo {author} {\bibfnamefont {J.}~\bibnamefont {Wrachtrup}},\
  }\bibfield  {title} {\bibinfo {title} {{Quantum Properties of Dichroic
  Silicon Vacancies in Silicon Carbide}},\ }\href@noop {} {\bibfield  {journal}
  {\bibinfo  {journal} {Physical Review Applied}\ }\textbf {\bibinfo {volume}
  {9}},\ \bibinfo {pages} {034022} (\bibinfo {year} {2018})}\BibitemShut
  {NoStop}%
\bibitem [{\citenamefont {Banks}\ \emph {et~al.}(2019)\citenamefont {Banks},
  \citenamefont {Soykal}, \citenamefont {Myers-Ward}, \citenamefont {Gaskill},
  \citenamefont {Reinecke},\ and\ \citenamefont {Carter}}]{Banks:2019je}%
  \BibitemOpen
  \bibfield  {author} {\bibinfo {author} {\bibfnamefont {H.~B.}\ \bibnamefont
  {Banks}}, \bibinfo {author} {\bibfnamefont {{\"O}.~O.}\ \bibnamefont
  {Soykal}}, \bibinfo {author} {\bibfnamefont {R.~L.}\ \bibnamefont
  {Myers-Ward}}, \bibinfo {author} {\bibfnamefont {D.~K.}\ \bibnamefont
  {Gaskill}}, \bibinfo {author} {\bibfnamefont {T.~L.}\ \bibnamefont
  {Reinecke}},\ and\ \bibinfo {author} {\bibfnamefont {S.~G.}\ \bibnamefont
  {Carter}},\ }\bibfield  {title} {\bibinfo {title} {{Resonant Optical Spin
  Initialization and Readout of Single Silicon Vacancies in 4H-SiC}},\
  }\href@noop {} {\bibfield  {journal} {\bibinfo  {journal} {Physical Review
  Applied}\ }\textbf {\bibinfo {volume} {11}},\ \bibinfo {pages} {024013}
  (\bibinfo {year} {2019})}\BibitemShut {NoStop}%
\bibitem [{\citenamefont {Iv{\'a}dy}\ \emph {et~al.}(2018)\citenamefont
  {Iv{\'a}dy}, \citenamefont {Abrikosov},\ and\ \citenamefont
  {Gali}}]{Ivady:2018cu}%
  \BibitemOpen
  \bibfield  {author} {\bibinfo {author} {\bibfnamefont {V.}~\bibnamefont
  {Iv{\'a}dy}}, \bibinfo {author} {\bibfnamefont {I.~A.}\ \bibnamefont
  {Abrikosov}},\ and\ \bibinfo {author} {\bibfnamefont {A.}~\bibnamefont
  {Gali}},\ }\bibfield  {title} {\bibinfo {title} {{First principles
  calculation of spin-related quantities for point defect qubit research}},\
  }\href@noop {} {\bibfield  {journal} {\bibinfo  {journal} {npj Computational
  Materials}\ }\textbf {\bibinfo {volume} {4}},\ \bibinfo {pages} {45}
  (\bibinfo {year} {2018})}\BibitemShut {NoStop}%
\bibitem [{\citenamefont {Bockstedte}\ \emph {et~al.}(2018)\citenamefont
  {Bockstedte}, \citenamefont {Sch{\"u}tz}, \citenamefont {Garratt},
  \citenamefont {Iv{\'a}dy},\ and\ \citenamefont {Gali}}]{Bockstedte:2018dh}%
  \BibitemOpen
  \bibfield  {author} {\bibinfo {author} {\bibfnamefont {M.}~\bibnamefont
  {Bockstedte}}, \bibinfo {author} {\bibfnamefont {F.}~\bibnamefont
  {Sch{\"u}tz}}, \bibinfo {author} {\bibfnamefont {T.}~\bibnamefont {Garratt}},
  \bibinfo {author} {\bibfnamefont {V.}~\bibnamefont {Iv{\'a}dy}},\ and\
  \bibinfo {author} {\bibfnamefont {A.}~\bibnamefont {Gali}},\ }\bibfield
  {title} {\bibinfo {title} {{Ab initio description of highly correlated states
  in defects for realizing quantum bits}},\ }\href@noop {} {\bibfield
  {journal} {\bibinfo  {journal} {npj Quantum Materials}\ }\textbf {\bibinfo
  {volume} {3}},\ \bibinfo {pages} {5645} (\bibinfo {year} {2018})}\BibitemShut
  {NoStop}%
\bibitem [{\citenamefont {Reshchikov}\ \emph {et~al.}(2011)\citenamefont
  {Reshchikov}, \citenamefont {Kvasov}, \citenamefont {Bishop}, \citenamefont
  {McMullen}, \citenamefont {Usikov}, \citenamefont {Soukhoveev},\ and\
  \citenamefont {Dmitriev}}]{Reshchikov:2011eo}%
  \BibitemOpen
  \bibfield  {author} {\bibinfo {author} {\bibfnamefont {M.~A.}\ \bibnamefont
  {Reshchikov}}, \bibinfo {author} {\bibfnamefont {A.~A.}\ \bibnamefont
  {Kvasov}}, \bibinfo {author} {\bibfnamefont {M.~F.}\ \bibnamefont {Bishop}},
  \bibinfo {author} {\bibfnamefont {T.}~\bibnamefont {McMullen}}, \bibinfo
  {author} {\bibfnamefont {A.}~\bibnamefont {Usikov}}, \bibinfo {author}
  {\bibfnamefont {V.}~\bibnamefont {Soukhoveev}},\ and\ \bibinfo {author}
  {\bibfnamefont {V.~A.}\ \bibnamefont {Dmitriev}},\ }\bibfield  {title}
  {\bibinfo {title} {{Tunable and abrupt thermal quenching of photoluminescence
  in high-resistivity Zn-doped GaN}},\ }\href@noop {} {\bibfield  {journal}
  {\bibinfo  {journal} {Physical Review B}\ }\textbf {\bibinfo {volume} {84}},\
  \bibinfo {pages} {075212} (\bibinfo {year} {2011})}\BibitemShut {NoStop}%
\bibitem [{\citenamefont {Kittel}(2004)}]{Kittel2004}%
  \BibitemOpen
  \bibfield  {author} {\bibinfo {author} {\bibfnamefont {C.}~\bibnamefont
  {Kittel}},\ }\href@noop {} {\emph {\bibinfo {title} {Introduction to Solid
  State Physics}}},\ \bibinfo {edition} {8th}\ ed.\ (\bibinfo  {publisher}
  {Wiley},\ \bibinfo {year} {2004})\BibitemShut {NoStop}%
\bibitem [{\citenamefont {Kresse}\ and\ \citenamefont {Hafner}(1993)}]{VASP}%
  \BibitemOpen
  \bibfield  {author} {\bibinfo {author} {\bibfnamefont {G.}~\bibnamefont
  {Kresse}}\ and\ \bibinfo {author} {\bibfnamefont {J.}~\bibnamefont
  {Hafner}},\ }\bibfield  {title} {\bibinfo {title} {{Ab Initio Molecular
  Dynamics For Liquid Metals}} } {\bibfield  {journal} {\bibinfo
  {journal} {Phys. Rev. B}\ }\textbf {\bibinfo {volume} {47}},\ \bibinfo
  {pages} {558} (\bibinfo {year} {1993})}\BibitemShut {NoStop}%
\bibitem [{\citenamefont {Perdew}\ \emph {et~al.}(2008)\citenamefont {Perdew},
  \citenamefont {Ruzsinszky}, \citenamefont {Csonka}, \citenamefont {Vydrov},
  \citenamefont {Scuseria}, \citenamefont {Constantin}, \citenamefont {Zhou},\
  and\ \citenamefont {Burke}}]{PBEsol}%
  \BibitemOpen
  \bibfield  {author} {\bibinfo {author} {\bibfnamefont {J.~P.}\ \bibnamefont
  {Perdew}}, \bibinfo {author} {\bibfnamefont {A.}~\bibnamefont {Ruzsinszky}},
  \bibinfo {author} {\bibfnamefont {G.~I.}\ \bibnamefont {Csonka}}, \bibinfo
  {author} {\bibfnamefont {O.~A.}\ \bibnamefont {Vydrov}}, \bibinfo {author}
  {\bibfnamefont {G.~E.}\ \bibnamefont {Scuseria}}, \bibinfo {author}
  {\bibfnamefont {L.~A.}\ \bibnamefont {Constantin}}, \bibinfo {author}
  {\bibfnamefont {X.}~\bibnamefont {Zhou}},\ and\ \bibinfo {author}
  {\bibfnamefont {K.}~\bibnamefont {Burke}},\ }\bibfield  {title} {\bibinfo
  {title} {{Restoring the Density-Gradient Expansion for Exchange in Solids and
  Surfaces}} }
  {\bibfield  {journal} {\bibinfo  {journal} {Phys. Rev. Lett.}\ }\textbf
  {\bibinfo {volume} {100}},\ \bibinfo {pages} {136406} (\bibinfo {year}
  {2008})}\BibitemShut {NoStop}%
\bibitem [{\citenamefont {Eriksson}\ \emph {et~al.}(2019)\citenamefont
  {Eriksson}, \citenamefont {Fransson},\ and\ \citenamefont
  {Erhart}}]{HIPHIVE}%
  \BibitemOpen
  \bibfield  {author} {\bibinfo {author} {\bibfnamefont {F.}~\bibnamefont
  {Eriksson}}, \bibinfo {author} {\bibfnamefont {E.}~\bibnamefont {Fransson}},\
  and\ \bibinfo {author} {\bibfnamefont {P.}~\bibnamefont {Erhart}},\
  }\bibfield  {title} {\bibinfo {title} {The hiphive package for the extraction
  of high-order force constants by machine learning} } {\bibfield  {journal} {\bibinfo
  {journal} {Advanced Theory and Simulations}\ }\textbf {\bibinfo {volume}
  {2}},\ \bibinfo {pages} {1800184} (\bibinfo {year} {2019})} \BibitemShut
  {NoStop}%
\bibitem [{\citenamefont {Togo}\ and\ \citenamefont {Tanaka}(2015)}]{PHONOPY}%
  \BibitemOpen
  \bibfield  {author} {\bibinfo {author} {\bibfnamefont {A.}~\bibnamefont
  {Togo}}\ and\ \bibinfo {author} {\bibfnamefont {I.}~\bibnamefont {Tanaka}},\
  }\bibfield  {title} {\bibinfo {title} {{First Principles Phonon Calculations
  In Materials Science}},\ }\href@noop {} {\bibfield  {journal} {\bibinfo
  {journal} {Scr. Mater.}\ }\textbf {\bibinfo {volume} {108}},\ \bibinfo
  {pages} {1} (\bibinfo {year} {2015})}\BibitemShut {NoStop}%
\bibitem [{\citenamefont {Krukau}\ \emph {et~al.}(2006)\citenamefont {Krukau},
  \citenamefont {Vydrov}, \citenamefont {Izmaylov},\ and\ \citenamefont
  {Scuseria}}]{hse06}%
  \BibitemOpen
  \bibfield  {author} {\bibinfo {author} {\bibfnamefont {A.~V.}\ \bibnamefont
  {Krukau}}, \bibinfo {author} {\bibfnamefont {O.~A.}\ \bibnamefont {Vydrov}},
  \bibinfo {author} {\bibfnamefont {A.~F.}\ \bibnamefont {Izmaylov}},\ and\
  \bibinfo {author} {\bibfnamefont {G.~E.}\ \bibnamefont {Scuseria}},\
  }\bibfield  {title} {\bibinfo {title} {Influence of the exchange screening
  parameter on the performance of screened hybrid functionals} } {\bibfield  {journal} {\bibinfo
  {journal} {The Journal of Chemical Physics}\ }\textbf {\bibinfo {volume}
  {125}},\ \bibinfo {pages} {224106} (\bibinfo {year} {2006})}\BibitemShut
  {NoStop}%
\bibitem [{\citenamefont {De\'ak}\ \emph {et~al.}(2011)\citenamefont {De\'ak},
  \citenamefont {Aradi},\ and\ \citenamefont {Frauenheim}}]{Deak2010PRB}%
  \BibitemOpen
  \bibfield  {author} {\bibinfo {author} {\bibfnamefont {P.}~\bibnamefont
  {De\'ak}}, \bibinfo {author} {\bibfnamefont {B.}~\bibnamefont {Aradi}},\ and\
  \bibinfo {author} {\bibfnamefont {T.}~\bibnamefont {Frauenheim}},\ }\bibfield
   {title} {\bibinfo {title} {Polaronic effects in TiO${}_{2}$ calculated by
  the HSE06 hybrid functional: Dopant passivation by carrier self-trapping} } {\bibfield  {journal}
  {\bibinfo  {journal} {Phys. Rev. B}\ }\textbf {\bibinfo {volume} {83}},\
  \bibinfo {pages} {155207} (\bibinfo {year} {2011})}\BibitemShut {NoStop}%
\bibitem [{\citenamefont {Lebedev}(1999)}]{Lebedev1999}%
  \BibitemOpen
  \bibfield  {author} {\bibinfo {author} {\bibfnamefont {A.~A.}\ \bibnamefont
  {Lebedev}},\ }\bibfield  {title} {\bibinfo {title} {Deep level centers in
  silicon carbide: A review} }
  {\bibfield  {journal} {\bibinfo  {journal} {Semiconductors}\ }\textbf
  {\bibinfo {volume} {33}},\ \bibinfo {pages} {107} (\bibinfo {year}
  {1999})}\BibitemShut {NoStop}%
\bibitem [{\citenamefont {MARKHAM}(1959)}]{MARKHAM1959}%
  \BibitemOpen
  \bibfield  {author} {\bibinfo {author} {\bibfnamefont {J.~J.}\ \bibnamefont
  {Markham}},\ }\bibfield  {title} {\bibinfo {title} {Interaction of normal
  modes with electron traps} } {\bibfield  {journal} {\bibinfo
  {journal} {Rev. Mod. Phys.}\ }\textbf {\bibinfo {volume} {31}},\ \bibinfo
  {pages} {956} (\bibinfo {year} {1959})}\BibitemShut {NoStop}%
\bibitem [{\citenamefont {Miyakawa}\ and\ \citenamefont
  {Dexter}(1970)}]{Miyakawa1970}%
  \BibitemOpen
  \bibfield  {author} {\bibinfo {author} {\bibfnamefont {T.}~\bibnamefont
  {Miyakawa}}\ and\ \bibinfo {author} {\bibfnamefont {D.~L.}\ \bibnamefont
  {Dexter}},\ }\bibfield  {title} {\bibinfo {title} {Phonon sidebands,
  multiphonon relaxation of excited states, and phonon-assisted energy transfer
  between ions in solids} }
  {\bibfield  {journal} {\bibinfo  {journal} {Phys. Rev. B}\ }\textbf {\bibinfo
  {volume} {1}},\ \bibinfo {pages} {2961} (\bibinfo {year} {1970})}\BibitemShut
  {NoStop}%
\bibitem [{\citenamefont {Falk}\ \emph {et~al.}(2013)\citenamefont {Falk},
  \citenamefont {Buckley}, \citenamefont {Calusine}, \citenamefont {Koehl},
  \citenamefont {Dobrovitski}, \citenamefont {Politi}, \citenamefont {Zorman},
  \citenamefont {Feng},\ and\ \citenamefont {Awschalom}}]{Falk:2013jq}%
  \BibitemOpen
  \bibfield  {author} {\bibinfo {author} {\bibfnamefont {A.~L.}\ \bibnamefont
  {Falk}}, \bibinfo {author} {\bibfnamefont {B.~B.}\ \bibnamefont {Buckley}},
  \bibinfo {author} {\bibfnamefont {G.}~\bibnamefont {Calusine}}, \bibinfo
  {author} {\bibfnamefont {W.~F.}\ \bibnamefont {Koehl}}, \bibinfo {author}
  {\bibfnamefont {V.~V.}\ \bibnamefont {Dobrovitski}}, \bibinfo {author}
  {\bibfnamefont {A.}~\bibnamefont {Politi}}, \bibinfo {author} {\bibfnamefont
  {C.~A.}\ \bibnamefont {Zorman}}, \bibinfo {author} {\bibfnamefont {P.~X.~L.}\
  \bibnamefont {Feng}},\ and\ \bibinfo {author} {\bibfnamefont {D.~D.}\
  \bibnamefont {Awschalom}},\ }\bibfield  {title} {\bibinfo {title} {{Polytype
  control of spin qubits in silicon carbide}},\ }\href@noop {} {\bibfield
  {journal} {\bibinfo  {journal} {Nature Communications}\ }\textbf {\bibinfo
  {volume} {4}},\ \bibinfo {pages} {1819} (\bibinfo {year} {2013})}\BibitemShut
  {NoStop}%
\bibitem [{\citenamefont {Zargaleh}\ \emph {et~al.}(2016)\citenamefont
  {Zargaleh}, \citenamefont {Eble}, \citenamefont {Hameau}, \citenamefont
  {Cantin}, \citenamefont {Legrand}, \citenamefont {Bernard}, \citenamefont
  {Margaillan}, \citenamefont {Lauret}, \citenamefont {Roch}, \citenamefont
  {von Bardeleben}, \citenamefont {Rauls}, \citenamefont {Gerstmann},\ and\
  \citenamefont {Treussart}}]{Zargaleh:2016kx}%
  \BibitemOpen
  \bibfield  {author} {\bibinfo {author} {\bibfnamefont {S.~A.}\ \bibnamefont
  {Zargaleh}}, \bibinfo {author} {\bibfnamefont {B.}~\bibnamefont {Eble}},
  \bibinfo {author} {\bibfnamefont {S.}~\bibnamefont {Hameau}}, \bibinfo
  {author} {\bibfnamefont {J.~L.}\ \bibnamefont {Cantin}}, \bibinfo {author}
  {\bibfnamefont {L.}~\bibnamefont {Legrand}}, \bibinfo {author} {\bibfnamefont
  {M.}~\bibnamefont {Bernard}}, \bibinfo {author} {\bibfnamefont
  {F.}~\bibnamefont {Margaillan}}, \bibinfo {author} {\bibfnamefont {J.~S.}\
  \bibnamefont {Lauret}}, \bibinfo {author} {\bibfnamefont {J.~F.}\
  \bibnamefont {Roch}}, \bibinfo {author} {\bibfnamefont {H.~J.}\ \bibnamefont
  {von Bardeleben}}, \bibinfo {author} {\bibfnamefont {E.}~\bibnamefont
  {Rauls}}, \bibinfo {author} {\bibfnamefont {U.}~\bibnamefont {Gerstmann}},\
  and\ \bibinfo {author} {\bibfnamefont {F.}~\bibnamefont {Treussart}},\
  }\bibfield  {title} {\bibinfo {title} {{Evidence for near-infrared
  photoluminescence of nitrogen vacancy centers in 4H-SiC}},\ }\href@noop {}
  {\bibfield  {journal} {\bibinfo  {journal} {Physical Review B}\ }\textbf
  {\bibinfo {volume} {94}},\ \bibinfo {pages} {86} (\bibinfo {year}
  {2016})}\BibitemShut {NoStop}%
\bibitem [{\citenamefont {Koehl}\ \emph {et~al.}(2017)\citenamefont {Koehl},
  \citenamefont {Diler}, \citenamefont {Whiteley}, \citenamefont {Bourassa},
  \citenamefont {Son}, \citenamefont {Janz{\'e}n},\ and\ \citenamefont
  {Awschalom}}]{Koehl:2017fd}%
  \BibitemOpen
  \bibfield  {author} {\bibinfo {author} {\bibfnamefont {W.~F.}\ \bibnamefont
  {Koehl}}, \bibinfo {author} {\bibfnamefont {B.}~\bibnamefont {Diler}},
  \bibinfo {author} {\bibfnamefont {S.~J.}\ \bibnamefont {Whiteley}}, \bibinfo
  {author} {\bibfnamefont {A.}~\bibnamefont {Bourassa}}, \bibinfo {author}
  {\bibfnamefont {N.~T.}\ \bibnamefont {Son}}, \bibinfo {author} {\bibfnamefont
  {E.}~\bibnamefont {Janz{\'e}n}},\ and\ \bibinfo {author} {\bibfnamefont
  {D.~D.}\ \bibnamefont {Awschalom}},\ }\bibfield  {title} {\bibinfo {title}
  {{Resonant optical spectroscopy and coherent control of Cr$^{4+}$ spin ensembles in
  SiC and GaN}},\ }\href@noop {} {\bibfield  {journal} {\bibinfo  {journal}
  {Physical Review B}\ }\textbf {\bibinfo {volume} {95}},\ \bibinfo {pages}
  {035207} (\bibinfo {year} {2017})}\BibitemShut {NoStop}%
\bibitem [{\citenamefont {Gottscholl}\ \emph {et~al.}(2019)\citenamefont
  {Gottscholl}, \citenamefont {Kianinia}, \citenamefont {Soltamov},
  \citenamefont {Bradac}, \citenamefont {Kasper}, \citenamefont {Krambrock},
  \citenamefont {Sperlich}, \citenamefont {Toth}, \citenamefont {Aharonovich},\
  and\ \citenamefont {Dyakonov}}]{Gottscholl:2019wd}%
  \BibitemOpen
  \bibfield  {author} {\bibinfo {author} {\bibfnamefont {A.}~\bibnamefont
  {Gottscholl}}, \bibinfo {author} {\bibfnamefont {M.}~\bibnamefont
  {Kianinia}}, \bibinfo {author} {\bibfnamefont {V.}~\bibnamefont {Soltamov}},
  \bibinfo {author} {\bibfnamefont {C.}~\bibnamefont {Bradac}}, \bibinfo
  {author} {\bibfnamefont {C.}~\bibnamefont {Kasper}}, \bibinfo {author}
  {\bibfnamefont {K.}~\bibnamefont {Krambrock}}, \bibinfo {author}
  {\bibfnamefont {A.}~\bibnamefont {Sperlich}}, \bibinfo {author}
  {\bibfnamefont {M.}~\bibnamefont {Toth}}, \bibinfo {author} {\bibfnamefont
  {I.}~\bibnamefont {Aharonovich}},\ and\ \bibinfo {author} {\bibfnamefont
  {V.}~\bibnamefont {Dyakonov}},\ }\bibfield  {title} {\bibinfo {title} {{Room
  Temperature Initialisation and Readout of Intrinsic Spin Defects in a Van der
  Waals Crystal}},\ }\href@noop {} {\bibfield  {journal} {\bibinfo  {journal}
  {{\tt arXiv:1906.03774}}} }\BibitemShut {NoStop}%
\end{thebibliography}

\providecommand{\noopsort}[1]{}\providecommand{\singleletter}[1]{#1}%

\end{document}